\documentclass[aps,jmp,amsmath,amssymb,reprint]{revtex4-1}
\usepackage{float}
\usepackage{graphicx}% Include figure files
\usepackage{dcolumn}% Align table columns on decimal point
\usepackage{bm}% bold math
\usepackage{graphicx}
\usepackage{subfigure}
\usepackage{physics}
\usepackage[english]{babel}
\usepackage{mathtools}
\usepackage[usenames, dvipsnames]{color}
\usepackage{hyperref}
\usepackage[normalem]{ulem}
\usepackage[utf8]{inputenc}
\usepackage{placeins}
\usepackage{graphicx}
\usepackage{amsmath}
\usepackage{makecell}
\usepackage[utf8]{inputenc}
\usepackage{amsmath,amssymb,amsfonts}
\usepackage{algorithmic}
\usepackage{float}
\usepackage{floatpag}
\usepackage[T1]{fontenc}
\begin{document}
\title{Insights into optical absorption and dark currents of the 6.1\AA\ Type-II superlattice absorbers for MWIR and SWIR applications}
%Choice of absorber and barrier for mid wavelength and long wavelength infrared photodetection }          
	\author{Anuja Singh}
	\author{Bhaskaran Muralidharan}
	\affiliation{Department of Electrical Engineering, Indian Institute of Technology Bombay, Powai, Mumbai-400076, India}
\date{\today}
\begin{abstract}
A holistic computational analysis is developed to calculate the quantum efficiency of InAs/GaSb superlattice-based photodetectors. Starting with the electronic band characteristics computed by taking the InSb/GaAs at the interface using the 8-band $\bf k.p$ approach, we demonstrate the impact of InAs and GaSb widths on the bandgap, carrier concentration, and the oscillator strength for type-II superlattice absorbers.  Subsequently, the alteration of these characteristics due to the extra AlSb layer in the M superlattice absorber is investigated.
Extending our models for determining TE- and TM-polarized optical absorption, our calculations reveal that the TE-polarized absorption shows a substantial influence near the conduction-heavy hole band transition energy, which eventually diminishes, owing to the dominant TM-contribution due to the conduction-light hole band transition. Extending our analysis to the dark currents, we focus mainly on Schokley-Read-Hall recombination and radiative recombination at lower temperatures, and show that Schokley-Read-Hall dominates at low-level injection. 
%Auger transitions are also predicted using the conduction mini-bandwidth and the heavy hole-light hole transition probabilities.
%For the dark current analysis, it is observed that the Schokley-Read-Hall recombination dominates at lower doping levels, whereas radiative recombination becomes dominant at higher doping levels.
%We predicted the Auger transitions using conduction mini-bandwidth and transition probabilities from heavy to light holes.
%Furthermore, a comprehensive comparison is conducted between the mid- and the long-wavelength type-II and M superlattice absorbers.
We show that short-wavelength and mid-wavelength M superlattice structures exhibit higher quantum efficiency than the corresponding same bandgap type-II superlattice with the lower diffusion dark current. 
%The long-wavelength M superlattice exhibits a longer diffusion length, due to its higher Auger and radiative recombination lifetimes, demonstrating clearly that it can be heavily doped in order to enhance quantum efficiency.
%Comparable characteristics are identified in the thinner M superlattice for mid-wavelength, resulting in a 7\% increase in quantum efficiency.
Further, we analyze the density of states blocked by the barrier; crucial for XBp photodetector after absorber examination. Our work thus sets a stage for a holistic and predictive theory aided analysis of the type-II superlattice absorbers, from the atomistic interfacial details all the way to the dark currents and absorption spectra. 
\end{abstract}
\maketitle
\section{\textbf{Introduction}}
Infrared photodetectors have various applications in the fields of light detection and ranging (LiDAR), detection of leakage of harmful gases and oil in industries, night vision cameras, defense, etc., \cite{plis2014inas,alshahrani2022emerging,rogalski2017inas,muller2020thermoelectrically} to name a few. Mercury cadmium telluride (HgCdTe) is a state-of-the-art material as it provides band tunability and higher electro-optical performance\cite{muller2020thermoelectrically,zhu2021high}. The substrate required for HgCdTe is the CdZnTe, which is expensive and such a ternary alloy has surface instabilities that limit its applications\cite{plis2014inas,mohseni1999uncooled}. Moreover, the permitted levels of mercury and tellurium are governed by the European Union's restrictions on hazardous substances (RoHS)\cite{muller2020thermoelectrically}.\\
\indent For higher operating temperature applications and environmental safety concerns, Antimony (Sb)-based type-II superlattice (T2SL) detectors are well-established and have proven to be a viable competitor of the HgCdTe- based devices \cite{zhu2021high,delmas2023high}. Superlattices based on InAs/GaSb are recognized for their tunable bandgap and higher effective masses. These properties lead to a photo-controlled absorption and lower dark currents in this material system\cite{maimon2006n,singh2022comprehensive,delmas2019comprehensive,boutramine2016electronic}. However, despite having promising dark current characteristics, the optical absorption, which is relatable to the electron-hole wavefunction overlaps\cite{lang2011interface}, i.e., the optical matrix elements for the type-II transition, needs enhancement because electrons and holes are spatially confined in the different layers. \cite{lang2013electronic,lang2011interface,singh2022comprehensive,mukherjee2021carrier}.\\
\indent In this work, we thoroughly investigate the InAs/GaSb/AlSb/GaSb based M superlattice (MSL) structures as viable candidates for the absorber, which has been demonstrated as a perfect barrier for the InAs/GaSb T2SL\cite{huang2011type,razeghi2010band,nguyen2010minority,zhu2022modeling}. However, their application as a potential absorber for the mid and short-wavelength region is still scarce in the literature. Prior to this work, a few studies evaluated the MSL for the short-wavelength application and have reported a minimal improvement in the absorption\cite{lang2013electronic,lang2011interface,singh2022comprehensive,razeghi2010band}.  \\
\indent Employing an in-depth analysis of the miniband effects on the transport properties, optical absorption, and recombination phenomenona, we explore the MSL absorber for both mid and short-wavelength infrared (MWIR and SWIR) detection \cite{mukherjee2021carrier,fang2022simulation,nghiem2017radiometric,muller2020thermoelectrically}. Previously, it has been reported that the MSL structures are adequate to provide improvements on the type-II transition\cite{razeghi2010band,lang2013electronic,ting2020inas}. In the MSL structure, an additional high bandgap AlSb layer inserted at the center of the GaSb layer splits the GaSb well into two-hole quantum wells, thereby shifting the center of the hole wavefunction near the interface, thereby increasing the absorption\cite{lang2013electronic,singh2022comprehensive,sharma2021emerging,ting2020inas}. However, the fact that the additional AlSb layer also increases the carrier confinement in the superlattice, which may lead to a reduction in absorption, was not explored \cite{lang2013electronic,ting2020inas,singh2022comprehensive}. Additionally, the MSL dark current phenomenon has not yet been investigated. Our work thereby fills this research gap via a thorough study of the optical and dark current properties of the same bandgap MWIR and SWIR T2SL and MSL.\\
 \indent We implement the 8-band $\bf k.p$ perturbation theory, which is well known for its accuracy near the Brillouin zone and accounts for the microscopic-interface effects (MIA) in this type-II broken aligned superlattices\cite{li2010intrinsic,galeriu2005k,mukherjee2021carrier}. Using the $\bf k.p$ model within the envelope function approximation (EFA) \cite{shulenberger2023electronic}, we calculate the miniband properties of these superlattices. The InSb and GaAs are taken as an interface to account for the lattice mismatch for this non-common atom superlattices\cite{delmas2019comprehensive,szmulowicz2006interfaces,alshahrani2023effect,livneh2012k}. The bandgap of T2SL and MSL and its variations with the composite material layer thickness are also calculated\cite{mukherjee2021carrier,singh2022comprehensive,livneh2012k}. Additionally, we mention the effect of localization on the bandgap and intrinsic carrier concentration\cite{klipstein2021type}. The Schokley-Read-Hall (SRH) electron and hole capture time constants are calculated to depict the carrier lifetime with the dopant concentrations\cite{hoglund2013influence,olson2015intensity,trupke2003temperature}. The electron and hole wave functions and their associated overlap integrals at the interface are also explored to determine the oscillator strength for the absorption\cite{brown2006short,dyksik2017electrical}. We also explore the vital SRH and radiative recombination phenomena that determine carrier diffusivity and carrier lifetime at lower temperature \cite{olson2015intensity,hoglund2013influence}. After depicting the electronic band properties and lifetimes, we employ Fermi's golden rule to calculate the transition energies participating in the absorption\cite{qiao2012electronic}. The transitions are depicted via the calculations of optical matrix elements for the conduction and valence subbands tarnsitions\cite{qiao2012electronic,wang2016enhancement,ahmed2016analysis,li2010intrinsic}. These matrix elements are also used to find the absorption coefficients and spontaneous emission rates of these superlattices \cite{wang2016enhancement,subashiev2004optical}. We also show the TE and TM absorption and photoluminescence spectra with incident photon energy\cite{wang2016enhancement,subashiev2004optical}.\\
\indent By analyzing the absorption spectrum, we report the dominance of TE absorption over TM near the conduction and heavy hole absorption edge, while at the energies above the corresponding conduction and light hole transitions energy, the TM absorption dominates \cite{wang2016enhancement,subashiev2004optical,trupke2003temperature}. We compute the photoluminescence spectra and report a red shift in bandgap and reduced radiative recombination rate while moving from SWIR to Long-wavelength(LWIR) region. In the case of MSL, we report the decrease in cut-off wavelength with the increase in the AlSb width and reduced full width at half maxima (FWHM) on increasing the AlSb width. 
%Further, the calculation of the miniband width of the bands and heavy hole-light hole split-off energy-like properties provide an in-depth understanding of the auger recombination, which limits the carrier lifetime at higher doping\cite{kitchin2000optical,grein1995long}. Radiative recombination is also found to be the dominant phenomenon over the SRH at higher doping concentrations, and the corresponding lifetime is also calculated \cite{qiao2012electronic,azarhoosh2016research}. 
Based on the above-described properties, we compare the same band gap T2SL and MSL for the MWIR and SWIR applications. The SWIR MSL is being found to have a higher quantum efficiency with similar diffusion dark current. Whereas, for the MWIR, the absorption increases in the MSL due to the increased carrier overlap at the interfaces, and the dark current is less. This makes MSL provide better electro-optical performance over T2SL.\\
\begin{figure*}[!htbp]
	\centering
		{\includegraphics[height=0.25\textwidth,width=0.8\textwidth]{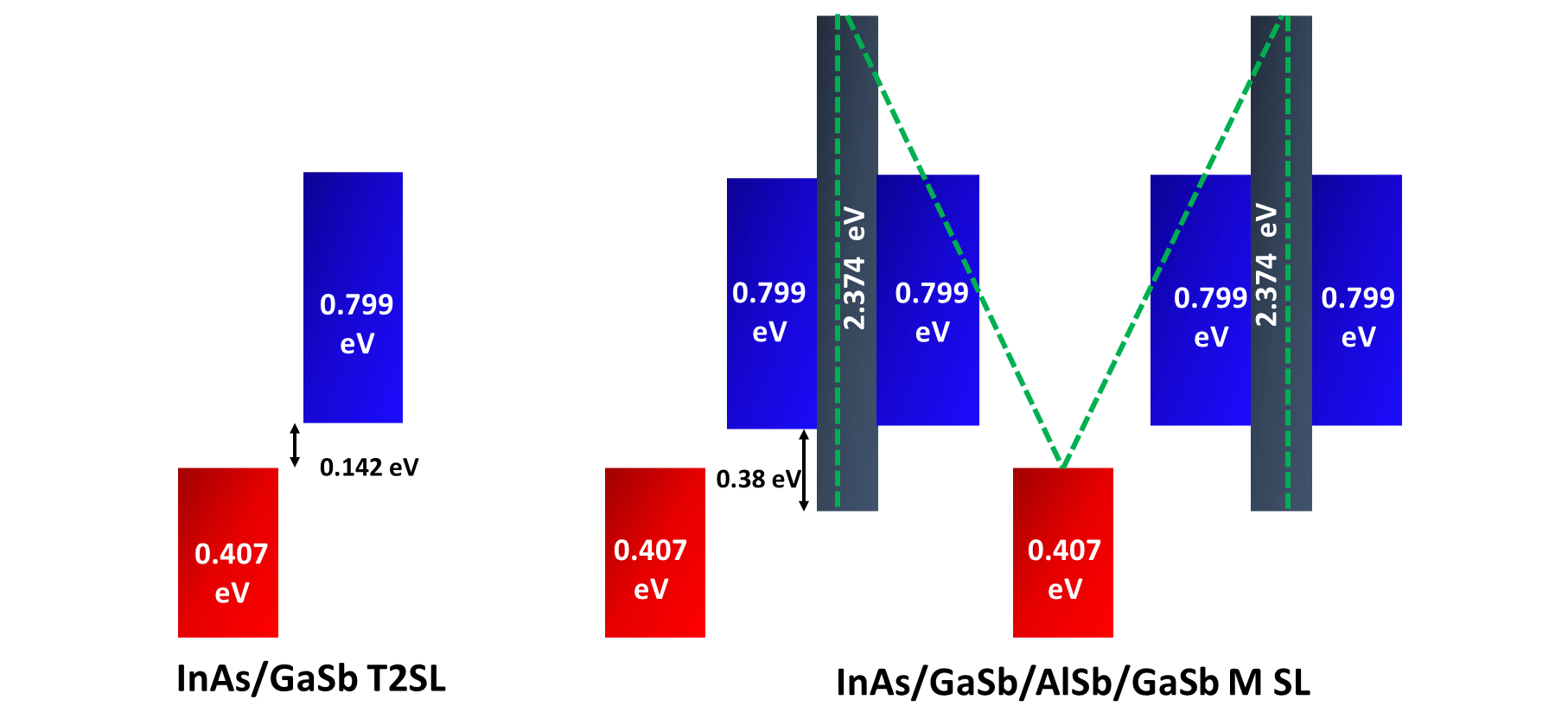}}
		\caption{Band alignment of T2SL and MSL. In MSL the high bandgap AlSb is inserted at the center of the GaSb, which divides the GaSb hole quantum well into two newly generated quantum wells.}
		\label{fig1}
\end{figure*}
\indent The rest of the paper is organized as follows. We outline the 8-band $\bf k.p$ method to calculate the energy dispersion for the InAs/GaSb and InAs/GaSb/AlSb/GaSb MSL. We discuss the approaches used for modeling the electronic band structure and absorption coefficient in Section II, which is further divided into three subsections Sec. II. A mentions the 8 band $\bf k.p$ model, Sec. II. B is for Fermi's golden rule formalism for the calculation of absorption and spontaneous emission spectra. In Sec. II.C., we describe the SRH and radiative lifetime. Further in Sec. III., we discuss the results of the findings and this is further divided into four subsections, which are Sec A.1 describes the T2SL absorber, Sec A.2 gives details of the MSL absorber, the later Sec A.3 and Sec A.4 compares the SWIR and MWIR T2SL and MSL absorbers, and in Sec. B, we provide the detailed design technique for the barriers, and in Sec. IV. we conclude this paper.
\section{Methodology}
\label{sec_theory}
\subsection{Electronic band structure by the 8 band $\bf{k.p}$ perturbation method}
\label{EK}
The 8-band $\bf k.p$ perturbation approach can precisely determine bandgap energies near the Brillouin zone while offering the correct interface treatment for the non-common atom type-II superlattices\cite{li2010intrinsic,galeriu2005k,singh2022comprehensive,szmulowicz2006interfaces}. For the type-II superlattice where there is a reduced Brillouin zone, the 8-band $\bf k.p$ method has been extensively utilized to determine the oscillator strength required to calculate the absorption coefficients and spontaneous emission rates\cite{qiao2012electronic}. In this research, we apply the 8-band coupled Hamiltonian—which encompasses the interaction of the conduction, heavy hole, light-hole, and split-off bands, as well as their double spins to calculate the band structure of the InAs/GaSb and InAs/GaSb/AlSb/GaSb superlattices\cite{qiao2012electronic}. The 8-band Lutinger-Kohn (LK) Hamiltonian is first decoupled into two 4x4 upper and lower Hamiltonians \cite{qiao2012electronic} which are represented as
\begin{equation} 
	H_{8x8}=\begin{bmatrix}
		H_{4x4}^{U}(k_t) & 0  \\ 0 & H_{4x4}^{L}(k_t)  \\
		
	\end{bmatrix},
\end{equation}
where, $k_{t}$ is the transverse wavenumber, $k_{t}=\sqrt{(k_{x}^2)+(k_{y}^2)}$.
The band energies and wavefunction are typically dependent on the magnitude of the $k_{t}$ and the azimuthal angle ($\phi$=$tan^{-1}(k_{y}/k_{x})$), the axial approximation is taken into the account\cite{qiao2012electronic}. The axial approximation simplifies the work that is required to perform the integration over the $k_t$ plane in the optical absorption calculations as it makes the energy subbands isotropic in the $k_t$ plane\cite{qiao2012electronic}. After the axial and basis transformation, the 4x4 upper and lower Hamiltonian can be represented as
\begin{equation} 
H_{4x4}^U=\\
\end{equation}
$\begin{bsmallmatrix}E_{c}+A&-\sqrt{3}V_{p}&-V_{p}+i\sqrt{2}U&-\sqrt{2}V_{p}-iU \\-\sqrt{3}V_{p}& Ev-P-Q & R_{p}+iS_{p} & \sqrt{2}R_{p}-i(S_{p})/\sqrt{2}  \\
  -V_{p}-i\sqrt{U}& R_{p}-iS_{p} & E_{v}-P+Q& -\sqrt{2}Q-i(S_{p})\sqrt{3/2}\\
		 -\sqrt{2}V_{p}+iU & \sqrt{2}R_{p}+i(S_{p})/\sqrt{2}& \sqrt{2}Q+i(S_{p})\sqrt{3/2}& E_{v}-P-\Delta\end{bsmallmatrix},$
\begin{equation} 
H_{4x4}^L=\\
\end{equation}
$\begin{bsmallmatrix}E_{c}+A&-\sqrt{3}V_{p}&-V_{p}+i\sqrt{2}U&-\sqrt{2}V_{p}-iU \\-\sqrt{3}V_{p}& E_{v}-P-Q & R_{p}+iS_{p} & \sqrt{2}R_{p}-i(S_{p})/\sqrt{2}  \\
  -V_{p}-i\sqrt{U}& R_{p}-iS_{p} & E_{v}-P+Q& -\sqrt{2}Q-i(S_{p})\sqrt{3/2}\\
		 -\sqrt{2}V_{p}+iU & \sqrt{2}R_{p}+i(S_{p})/\sqrt{2}& \sqrt{2}Q+i(S_{p})\sqrt{3/2}& E_{v}-P-\Delta\\
   \end{bsmallmatrix},$\\

where,  $E_{c}$ and $E_{v}$ are the conduction and valence band edges, respectively. The interband mixing parameter or Kane's parameter is represented as  $P_{cv}$. The $\gamma_1$, $\gamma_2$, and $\gamma_3$ are the modified Luttinger parameters and $m_{c}'$ is the corrected effective mass. The $k_{z}$ is replaced by the $-i\partial/\partial z$ operator in the above equations to construct the  Hamiltonian. The other parameters are given by 
\begin{equation} 
A=\frac{\hbar^2}{2m_c'}(k_{t}^2+k_{z}^{2}),
\end{equation}
\begin{equation} 
P=\gamma_{1}\frac{\hbar^2}{2m_0}(k_{t}^2+k_{z}^{2}),
\end{equation}
\begin{equation} 
Q=\gamma_{2}\frac{\hbar^2}{2m_0}(k_{t}^2-2k_{z}^{2}),
\end{equation}
\begin{equation} 
R_{p}=-\frac{\hbar^2}{2m_0}\sqrt{3}\frac{(\gamma_{1}+\gamma_{3})}{2}(k_{t}^2),
\end{equation}
\begin{equation} 
S_{p}=-\frac{\hbar^2}{2m_0}2\sqrt{3}\gamma_{3}(k_{t}*k_{z}),
\end{equation}
\begin{equation} 
V_{p}=-\frac{1}{\sqrt{6}}P_{cv}k_{t},
\end{equation}
\begin{equation} 
U=-\frac{1}{\sqrt{3}}P_{cv}k_{z}.
\end{equation}
 In these superlattices, as the additional potential due to the band, alignment is superimposed over the underlying atomic potentials, we implement the 8-band $\bf k.p$ method with the envelope function approximation (EFA). To obtain the energies and their corresponding wavefunctions, the finite difference method (FDM) with periodic boundary conditions are employed to solve this Hamiltonian\cite{galeriu2005k,mukherjee2021carrier,seyedein2023band}.
Additionally, we treat the interface as an InSb layer with 3\AA width in one superlattice period to account for the lattice mismatch in these non-common atom configurations\cite{szmulowicz2006interfaces,delmas2019comprehensive}.
The interband momentum matrix parameter is calculated by the weighted average of InAs, GaSb, InSb, and AlSb layers. The parameters for InAs, GaSb, InSb, and AlSb materials are provided in Table \ref{table1}. Figure. \ref{fig1} shows the band alignment in T2SL and MSL. The following equations are implemented to calculate the density of states of these superlattices utilizing the 8-band $\bf k.p$ method \cite{benchtaber2020correlation,benaadad2021long},
%In order to calculate the density-of-states of these superlattices by utilizing the 8 band k.p method following equations were implemented\cite{benchtaber2020correlation,benaadad2021long},
\begin{equation}
\left\{ \begin{aligned} 
  DOS^{i}(E)\ &=\frac{m^{*}}{\pi^{2}\hbar^{2}}k_{z}(E) \ E_{min}^{i}\leq E\leq E_{max}^{i},\\
  &= 0  \ otherwise,
\end{aligned} \right.
\label{alpha}
\end{equation}
where, i is the index that represents the band, $E_{min}$ and $E_{max}$ are the energies at the $k_{z}=0$, and  $k_{z}=\pi/L$, respectively, $m^{*}$ is the effective mass, and $k_{z}$ is the wavevector along the growth direction. The miniband width of the superlattice is defined by the term $Emax^{i}-Emin^{i}$.
%$m^{*}$ is the effective mass, and $k_{z}$ is the wavevector along the growth direction.
\begin{table}[ht]
\addtolength{\tabcolsep}{-1pt}
\small
\footnotesize
\caption{Material parameters to calculate the electronic band structure by the 8 band $\bf k.p$ perturbation method at $T=77K$.} 
\label{table1}
\begin{center}    

\begin{tabular}{|l|l|l|l|l|l|}

\hline
\rule[-1ex]{0pt}{2.5ex} 	Parameters & InAs & GaSb & AlSb & InSb  \\
\hline
\rule[-1ex]{0pt}{2.5ex}   \makecell{Valence band offset \\  (VBO) [$eV$] }  & -0.56 & 0 & -0.38& 0.03\\

\hline

\rule[-1ex]{0pt}{2.5ex}   \makecell{Lattice constant \\(\AA) } 	& 6.0522 & 6.0854 & 6.1297 & 6.4717\\
\hline

\rule[-1ex]{0pt}{2.5ex} \makecell{Spin-orbit splitting \\(SO) [$eV$] } & 0.38& 0.75 & 0.65 &0.81\\
\hline
\rule[-1ex]{0pt}{2.5ex} \makecell{Optical matrix parameter \\( Ep [$eV$]) } & 21.5 & 22.4 &18.7 &23.3\\
\hline
\rule[-1ex]{0pt}{2.5ex}  	\makecell{Luttinger parameter \\($\gamma_{1}$) } & 19.4& 11.84&4.15 &32.4\\
\hline
\rule[-1ex]{0pt}{2.5ex}  	\makecell{Luttinger parameter \\($\gamma_{2}$) } & 8.545 & 4.25&1.28& 13.3\\  
\hline
\rule[-1ex]{0pt}{2.5ex}  	\makecell{ Luttinger parameter \\($\gamma_{3}$) } & 9.17 & 5.01&1.75 &15.15\\  
\hline
\rule[-1ex]{0pt}{2.5ex}  	\makecell{Corrected Luttinger parameter \\($\gamma1$) }  & 6.30& 4.86&1.76 &16.78 \\
\hline
\rule[-1ex]{0pt}{2.5ex} \makecell{Corrected Luttinger parameter \\($\gamma2$) } & 1.99 & 0.76&0.085 & 5.61\\  
\hline
\rule[-1ex]{0pt}{2.5ex} \makecell{Corrected Luttinger parameter \\($\gamma3$) }  & 2.62 & 1.52&0.55 & 7.46\\  
\hline
\rule[-1ex]{0pt}{2.5ex} \makecell{Effective mass electron\\($m_e/m_{0}$)}& 0.022&0.0412&0.13 &0.0135\\
\hline
\rule[-1ex]{0pt}{2.5ex}  \makecell{Corrected Effective mass electron\\($m_e^{'}/m_{0}$)}& 2.51&1.053&2.716&1.84 \\
\hline
\rule[-1ex]{0pt}{2.5ex} 	 \makecell{Energy band gap \\at T=77K($eV$)}& 0.407 & 0.799&2.374 &0.227 \\
\hline 
\end{tabular}
\end{center}
\end{table} 
\subsection{Absorption coefficient and optical moment matrix element}
\label{absorption}
The optical absorption coefficient represents the value that how far the light travels in the material before being absorbed. It gives an idea of how readily the light is absorbed in the material and contributes to the generation of electron-hole pairs, which will later contribute to the photocurrent in the device. It is a product of the joint densities of states and the Fermi function, which can be calculated by Fermi's Golden Rule \cite{qiao2012electronic} and represented as
\begin{equation}
\begin{split}
\alpha(\hbar w)=\frac{\pi e^{2}}{(n_{r}c \epsilon m_{0}^{2}w)}\sum_{\sigma_{1},\sigma_{2}}^{U,L}\sum_{n,m}\frac {1}{L_{T}} \int_{0}^{2\pi}\frac{d\phi}{2\pi}\int_{0}^{\infty}\frac{d\phi}{2\pi}\frac{k_{t}dk_{t}}{2\pi}\\
|<\psi_{c}^{\sigma_{1},n}|e.p|\psi_{v}^{\sigma_{2},m}>|^{2}(f_{v}^{\sigma_{2,m}}(k_{t})-f_{c}^{\sigma_{1,n}}(k_{t})L(k_{t},\hbar*w),
\label{ab}
\end{split}
\end{equation}
where n represents the refractive index of the superlattice, $\epsilon$ denotes the permittivity of the free space, $E_{c}(k_{t}), E_{v}(k_{t})$, are conduction and valence band energies calculated by the 8 band $\bf k.p$ theory at a particular $k_{t}$, m and n are the valence and conduction band indexes, w is the incident photon frequency, and $f_{c},f_{v}$ are the Fermi functions and $E_{fn}, 
 E_{fp}$, are the quasi-Fermi levels calculated at certain doping, and $L$ is the Lorentzian function, given by
\begin{equation}
L(k_{t},\hbar w)=\frac{1}{\gamma \sqrt{2\pi}}exp{\frac{-({E_{c}(k_{t})-E_{v}(k_{t})-\hbar w})^{2}}{2\gamma^{2}}},
\end{equation}
where the $\gamma$ parameter is used for the scattering processes, also called as broadening function and the value taken in this work is $9meV$ both for the T2SL and MSL.  The Fermi functions are denoted as $f_{c}$ and $f_{v}$ and can be expressed as
\begin{equation}
f_{c}^{\sigma_{1},n}(k_{t})=\frac{1}{1+e^\frac{{E_c(K_{t})-E_{fn}}}{k_{B}T}},
\label{fc}
\end{equation}
\begin{equation}
f_{v}^{\sigma_{2},m}(k_{t})=\frac{1}{1+e^\frac{{E_v(K_{t})-E_{fp}}}{k_{B}T}}
\label{fv}
\end{equation}
where, $k_{B}$ is the Boltzmann constant, and other parameters are the same as in the functions discussed above.\\
\indent Further, the term in equation \ref{ab}, i.e., $|<\psi_{c}^{\sigma_{1},n}|e.p|\psi_{v}^{\sigma_{2},m}>|$ is the optical matrix moment elements which represents the transition probability from the filled valence band to the empty conduction band and is calculated by equations given in \cite{qiao2012electronic}, which mainly depends on the overlap of the envelope wavefunction w.r.t to the space and the transverse wavevector. Both the TE and TM polarization matrix elements which are required for the calculation of the absorption coefficients are calculated by the equation as given in \cite{qiao2012electronic}. Where, the absorption includes the upward transition, the opposite of absorption is spontaneous emission, which considers only the downward transition and requires the initial conduction band state to be occupied, and the final valence band state to be empty. The spontaneous emission rate is given by
\begin{equation}
\begin{split}
e_{spon}(\hbar w)=\frac{n_{r}e^{2}w}{c^{3}\pi\hbar\epsilon m_{0}^{2}}\sum_{\sigma_{1},\sigma_{2}}^{U,L}\sum_{n,m}\frac {1}{L_{T}} \int_{0}^{2\pi}\frac{d\phi}{2\pi}\int_{0}^{\infty}\frac{d\phi}{2\pi}\frac{k_{t}dk_{t}}{2\pi}\\
|<\psi_{c}^{\sigma_{1},n}|e.p|\psi_{v}^{\sigma_{2},m}>|^{2}(f_{c}({k_{t}})^{\sigma_{1},n})(1-f_{v}({k_{t}})^{\sigma_{2},m})\\L(k_{t},\hbar*w)),
\label{em}
\end{split}
\end{equation}
where all the terms are similar to absorption but according to the occupancy of electrons and holes in conduction and valence bands, there is a difference in the terms representing the Fermi function. 
\subsection{Carrier lifetime}
\label{lifetime}
The generated electron-hole pairs in the active region should get transmitted to their respective contacts for the photocurrent. In addition, the diffusion length of the minority carriers are to be higher than the absorption length only then they can participate in the photoconduction, otherwise, they will just lost by the recombination process in the device. Therefore, the minority carrier's lifetime is a crucial parameter, as it determines both the dark current and maximum operating temperature for the acceptable performance of the photodetectors. The diffusion-limited detector dark current \cite{olson2015intensity} in the n-type doped material is given by
\begin{equation}
J_{diff}=q\frac{ni^{2}t}{n_{0}*\tau_{mc}},
\label{jd}
\end{equation}
where $q$ represents the electronic charge, $n_{i}$ is the intrinsic carrier density of the material, $n_{0}$ is the electron concentration which is in the majority, and $t$ is the thickness of the active region, and $\tau_{mc}$ is the minority carrier lifetime. The lifetime of the carriers depends on the various recombination phenomena, which are Shockley-Read-Hall(SRH), radiative, and Auger recombination. 
The SRH recombination process is due to the defect levels present in the material, which restricts their flow as the carriers get trapped in the trap level. The minority SRH lifetime\cite{blakemore2002semiconductor,bandara2011doping} due to an electron in the p-type absorber is given by
\begin{equation}
\tau(SRH)_{n}=\frac{\tau_{n_{0}}(p_{0}+p_{1}+\Delta p)+\tau_{p_{0}}(n_{0}+n_{1}+\Delta n)}{p_{0}+n_{0}+\Delta n},
\end{equation}
where, $\Delta n$ and $\Delta n $ are the excess carrier concentration($\Delta n$=$\Delta p$) and the $n_{1}$ and $p_{1}$ can be expressed as
\begin{equation}
n_{1}=n_{i}exp(\frac{E_{t}-E_{i}}{k_{B}T}),
p_{1}=n_{i}exp(\frac{E_{i}-E_{t}}{k_{B}T}),
\end{equation}
where $E_{i}$ is the intrinsic energy level, and  the $E_{t}$ is the trap level, respectively. The SRH lifetime \cite{hoglund2013influence} depends on the electron and hole capture constants which are given by
\begin{equation}
\tau_{n0}=\frac{1}{\sigma_{n}v_{n}N_{t}},
\tau_{p0}=\frac{1}{\sigma_{p}v_{p}N_{t}},
\label{tau}
\end{equation}
where, $\sigma_{n,p}$ are the capture cross-section area for the electron and hole defect, respectively, and $v_{p,n}$ is the thermal velocity of the carriers.\\
%For an n-type material and assuming the trap level at midgap, the SRH lifetime \cite{olson2015intensity} can be represented as
%\begin{equation}
%\tau_{SRH}^{-1}=\frac{n_{0}+\Delta n}{\tau_{p_{0}}(n_{0}+\Delta n)+\tau_{n_{0}}(\Delta n)}.
%\label{taulife}
%\end{equation}
%In case of low level injection, with the $\Delta n <<n_{0}$, $\tau_{SRH}$ can be denoted as $\tau_{p_{0}}$ and similarly for the p-type it will be $\tau_{n_{0}}$.\\
The radiative recombination coefficient which involves the band-band transitioning from the conduction band to the valence band \cite{trupke2003temperature,azarhoosh2016research} is calculated by the following equation
%\begin{equation}
%B_{rad}(T)=\frac{1}{n_{i}^2\pi^2\hbar^3 c^{2}} \int_{0}^\infty n_{r}^2 (\hbar w)^{2}\alpha(\hbar w,T)e^{(\frac{-\hbar w}{k_{B}T})}d(\hbar w),
%\label{rad}
%\end{equation}
where $\alpha(\hbar w, T)$ is the absorption coefficient at frequency w and temperature T, and $n_{r}$ is the refractive index of the superlattice material.
\FloatBarrier
\section{Results and discussions}
Electrons in the T2SL are delocalized, which leads to conduction via minibands, while hopping occurs for holes due to their higher localization in GaSb wells \cite{klipstein2021type,singh2022comprehensive,hussain2022electronic,ting2011superlattice}. These localization effects depend on the thickness of InAs and GaSb taken in one superlattice period. The carriers' localization and wavefunction overlap play a significant role in determining the optical and dark current characteristics of T2SL\cite{nghiem2017radiometric,imbert2015electronic,singh2022quantum}. For the design of barrier-based photodetectors (XBp), three regions in particular—contact(X), barrier(B), and absorber(p-type)—must be considered \cite{nguyen2009minority}.
%In particular, three regions of contact, barrier, and absorber must be envisaged for the design of photodetectors \cite{nguyen2009minority}.
T2SL is applicable to design these regions due to the tunability of the conduction and valence band edges\cite{akhavan2016superlattice,mukherjee2018improved,delmas2019comprehensive,razeghi2010band}.
The valence and conduction band offsets between the barrier and absorber layers decide the flow of photogenerated carriers from the absorber to the contacts\cite{kopytko2015engineering,li2019dark}. In this work, we examine the effect of carrier localization on the absorption and dark current properties of T2SL and MSL for the SWIR and MWIR applications. The same bandgap T2SL and MSL are also explored to study the effect of InAs, GaSb, and AlSb layer widths on quantum efficiency and minority carrier lifetime. We also showcase the design of the barrier to design the XBp detectors. The barrier blocked certain density-of-states which restricts the photogenerated carriers' flow in one direction only.

\begin{figure}[!htbp]
	\centering
		{\includegraphics[height=0.25\textwidth,width=0.5\textwidth]{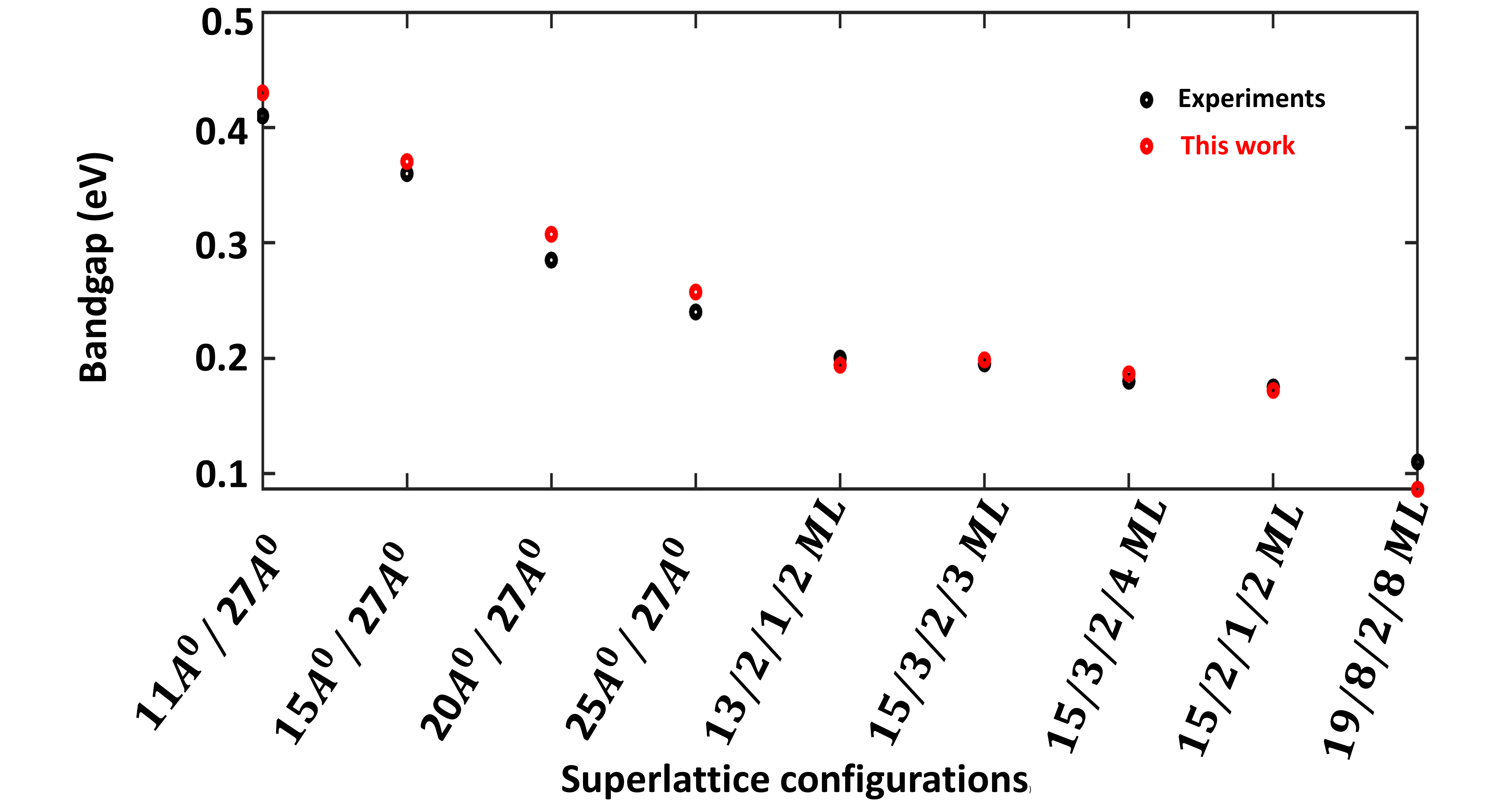}}

			\caption{Calculated bandgaps for T2SL and MSL by the 8 band $\bf k.p$ method. The calculated values are in close proximity to the experimentally observed values \cite{lang2013electronic}. }
		\label{fig2}
\end{figure}
\label{sec_res}
\begin{figure}[!htbp]
	\centering
{\includegraphics[height=0.25\textwidth,width=0.5\textwidth]{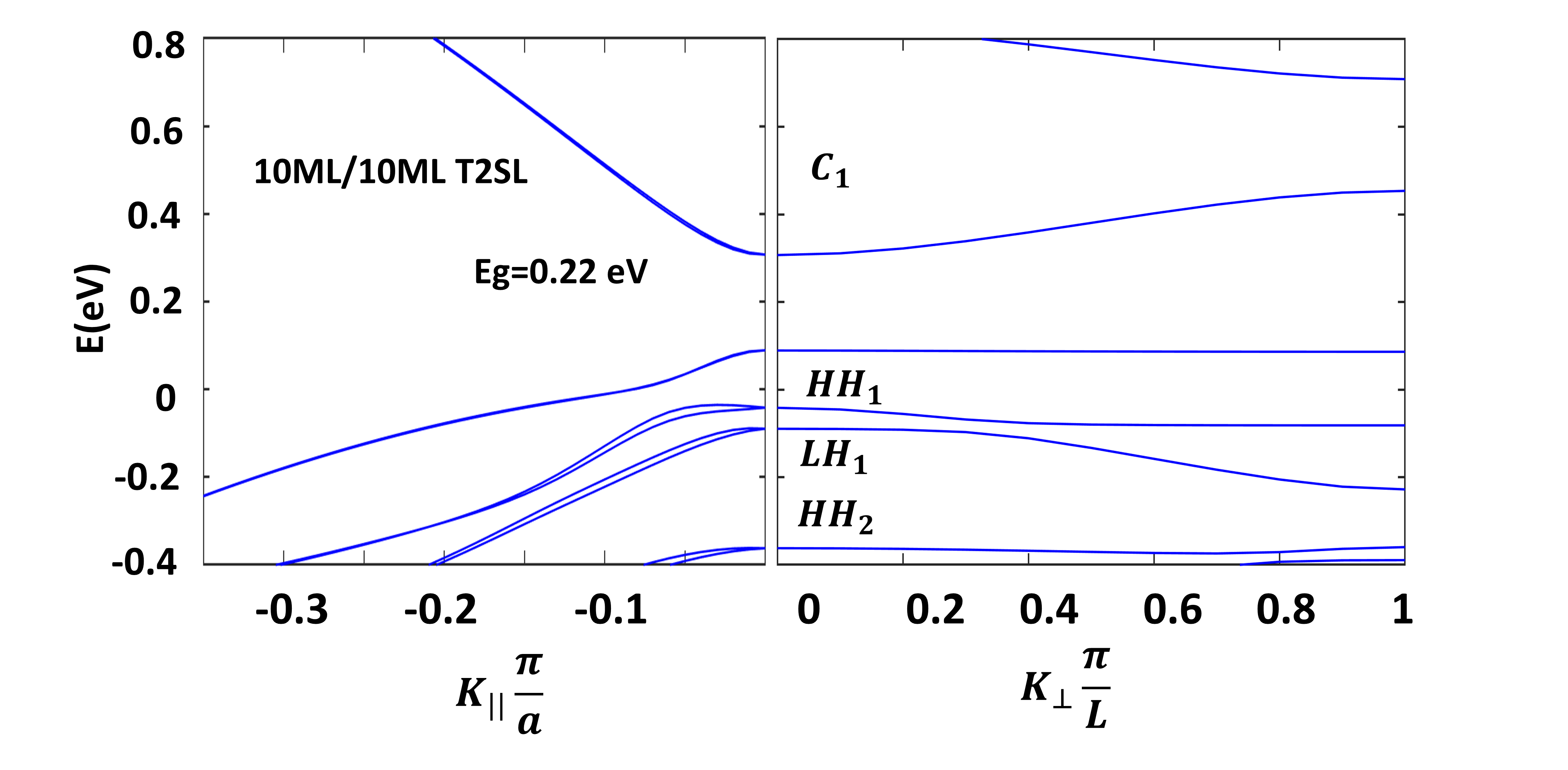}}
			\caption{Energy dispersion plot for the 10ML/10ML superlattice both in parallel and out of the plane directions calculated by the 8 band $\bf k.p$ method by taking the InSb(GaAs) at InAs/GaSb(GaSb/InAs) interface. The difference in energy at $k_{\perp}=0$ and $k_{\perp}=\pi/L$ in the out-of-plane E-K dispersion plot is the miniband width of the T2SL. The observed miniband width is 146 $meV$. The split in the valences bands are due to the interface. Also, there is an anti-crossing between the light hole and heavy hole bands in the perpendicular direction due to interface consideration in the band structure calculations.}
		\label{fig3}
\end{figure}
\begin{figure*}[!htbp]
	\centering
		{\includegraphics[height=0.35\textwidth,width=1\textwidth]{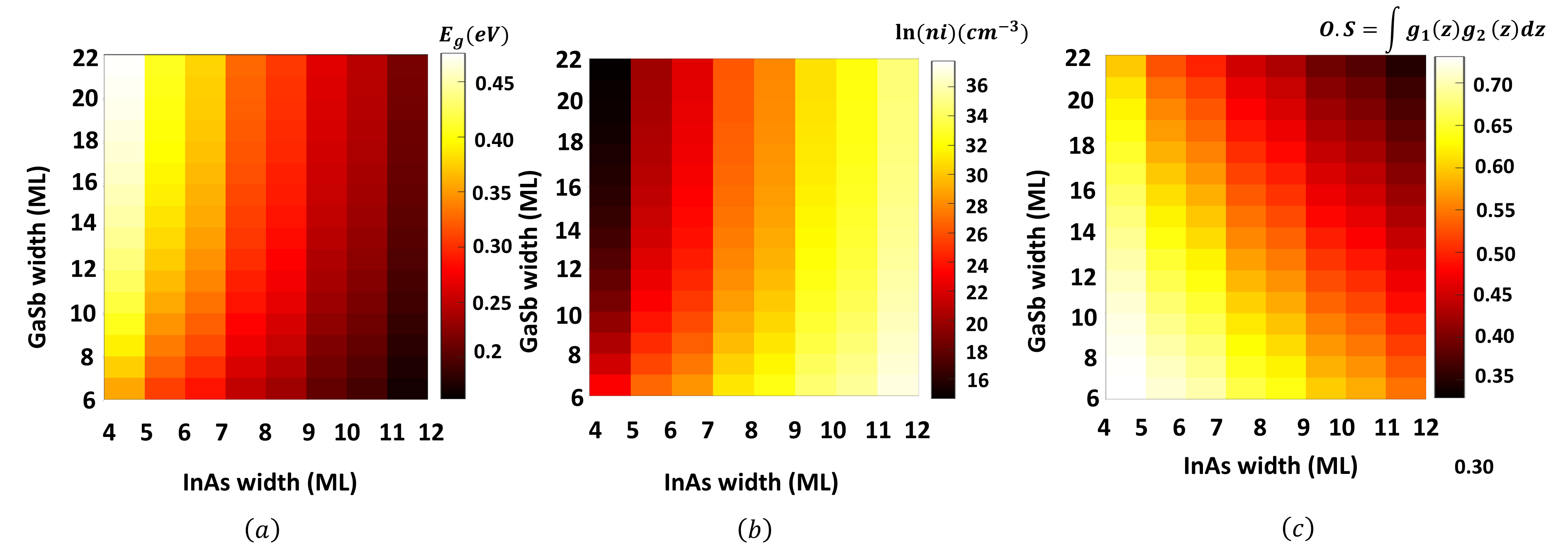}}
		\caption{The bandgap, intrinsic carrier concentration, and oscillator strength of T2SL with the various widths of InAs and GaSb are calculated by 8 bands $\bf k.p$ perturbation theory at T=77K. (a) Bandgap (b) Intrinsic carrier concentration ($n_{i}$)  is calculated by using the effective masses and bandgaps obtained by the band structure. (c) The oscillator strength (O.S) is calculated by the integration of the first conduction and first heavy hole band in the growth axis at $k_{t}=0$. It can be seen as the InAs width is increasing the bandgap is reducing sharply. The $n_{i}$ increases as the InAs width increases, it is due to the fact that as the InAs width increases, the effective mass of the hole increases, as well as the bandgap reduces. At a constant InAs width, while increasing the GaSb width, the bandgap increases, and hence there is a reduction in the $n_{i}$. The oscillator strength decreases as the widths of InAs and GaSb increase. As the localization increases, the overlap between the C1-HH1 decreases, as the carriers are more localized in their wells and interact less with the carriers in the consecutive wells.}
		\label{fig4}
\end{figure*}
\subsection{Absorber}
 \indent Reduced SRH, Auger, and radiative recombination dark currents that led to higher operating temperature can be achieved owing to the spatial confinement of carriers in the InAs/GaSb T2SL(holes in GaSb well and electrons in InAs well)\cite{kopytko2015engineering,li2019dark,downs2013progress}. 
 In this work, we want to design an absorber for the XBp barrier-based IR detectors, where most of the depletion region falls in the barrier(B) region, and due to its higher bandgap, the SRH dark current component reduces. The barrier further reduces the band-to-band dark tunneling currents\cite{li2019dark,singh2022comprehensive,kopytko2015engineering}. To achieve higher quantum efficiency, we investigate a p-type doped absorption region almost at equilibrium (the depletion region shifts more toward the barrier region).
 %For the MWIR and LWIR applications, we examine the T2SL and MSL absorbers. We further compare the optical and dark current characteristics of these absorbers and showcase the barriers' design and the blocked density-of-states by the barriers which only restrict their flow in one direction.
\subsubsection{\textbf{InAs/GaSb Type-II superlattice Absorber}}
\indent 
 To obtain the electronic band structure characteristics of T2SL, we apply an 8-band $\bf k.p$ perturbative method as discussed in section II.A. We consider InSb and GaAs as an alternative interface between InAs and GaSb, and the width of both InSb and GaAs is taken as 1.5\AA\cite{qiao2012electronic,delmas2019comprehensive}. We plot our simulated results with existing experimental or simulation values from the literature in Fig. \ref{fig2} to validate our results. We then calculate the band structure of 10ML/10ML T2SL, as shown in Fig.\ref{fig3}, where we see an almost dispersion-less E-k plot in the growth direction for holes because they are highly localized in the GaSb quantum well. While the electrons are delocalized, this can be seen from both the energy dispersion and the conduction band mini-bandwidth  (difference between energy at $k_{\perp}$=0 and $k_{\perp}=\pi/L$, the mini-bandwidth of the holes are a few $meV$, while an electron's miniband width is around $146 meV$). The heavy-hole, light-hole band splitting is there due to the interface consideration.\\

\begin{figure}[!htbp]
	\centering
 {\includegraphics[height=0.25\textwidth,width=0.35\textwidth]{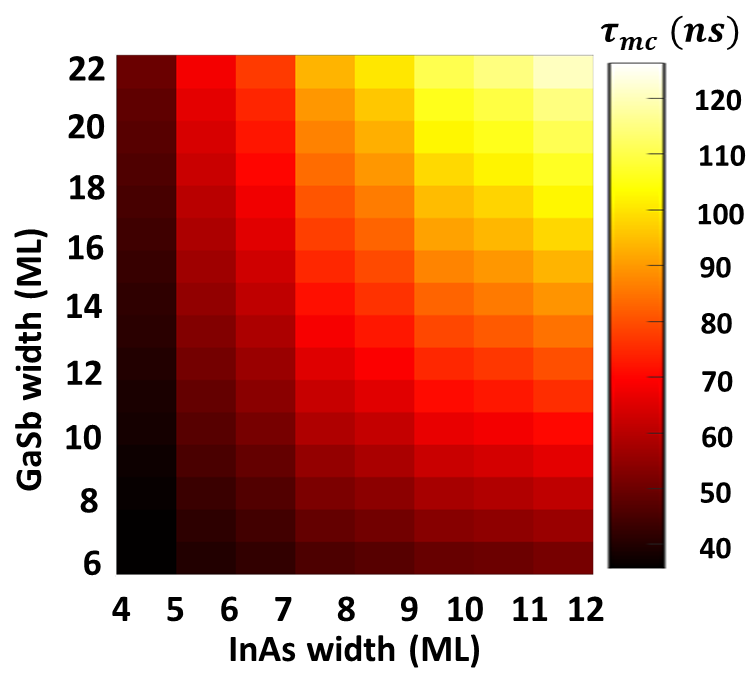}}
		\caption {SRH minority carrier lifetime for the varies widths of InAs and GaSb are calculated by the 8 band k.p perturbation theory at T=77K.}
		\label{fig5}
\end{figure}

\begin{figure}[!htbp]
	\centering
		{\includegraphics[height=0.23\textwidth,width=0.45\textwidth]{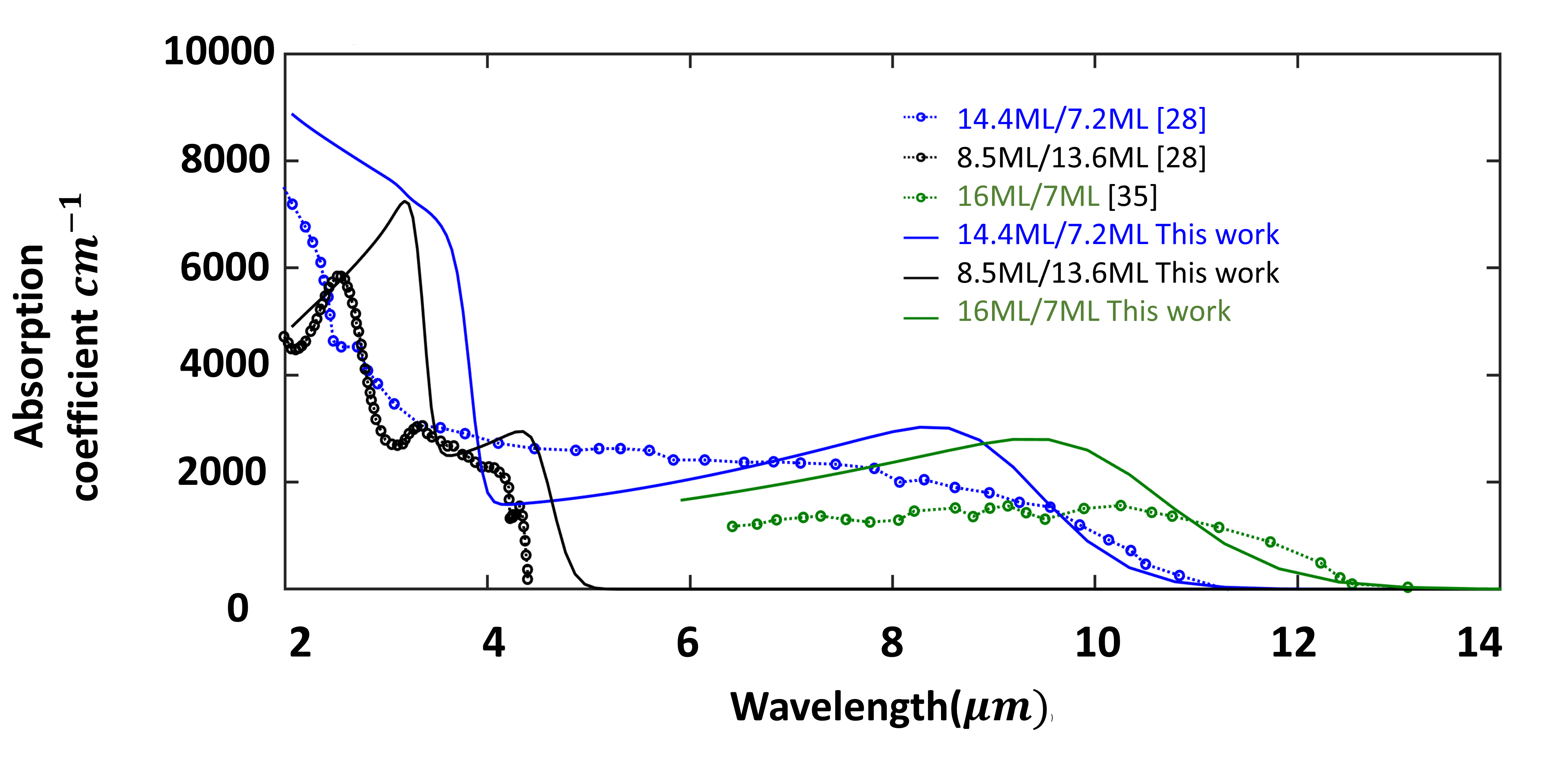}}

			\caption{Calculated Absorption coefficient for T2SL by utilizing the 8 band k.p method and Fermi's golden rule for three T2SL configurations. T2SL1: 8.5ML/13.6ML, T2SL2: 14.4ML/7ML, T2SL3:16ML/7ML. The calculated values are in close proximity to the experimentally observed values \cite{livneh2012k,qiao2012electronic}. }
		\label{fig6}
\end{figure}
\begin{figure}[!htbp]
	\centering

{\includegraphics[height=0.22\textwidth,width=0.45\textwidth]{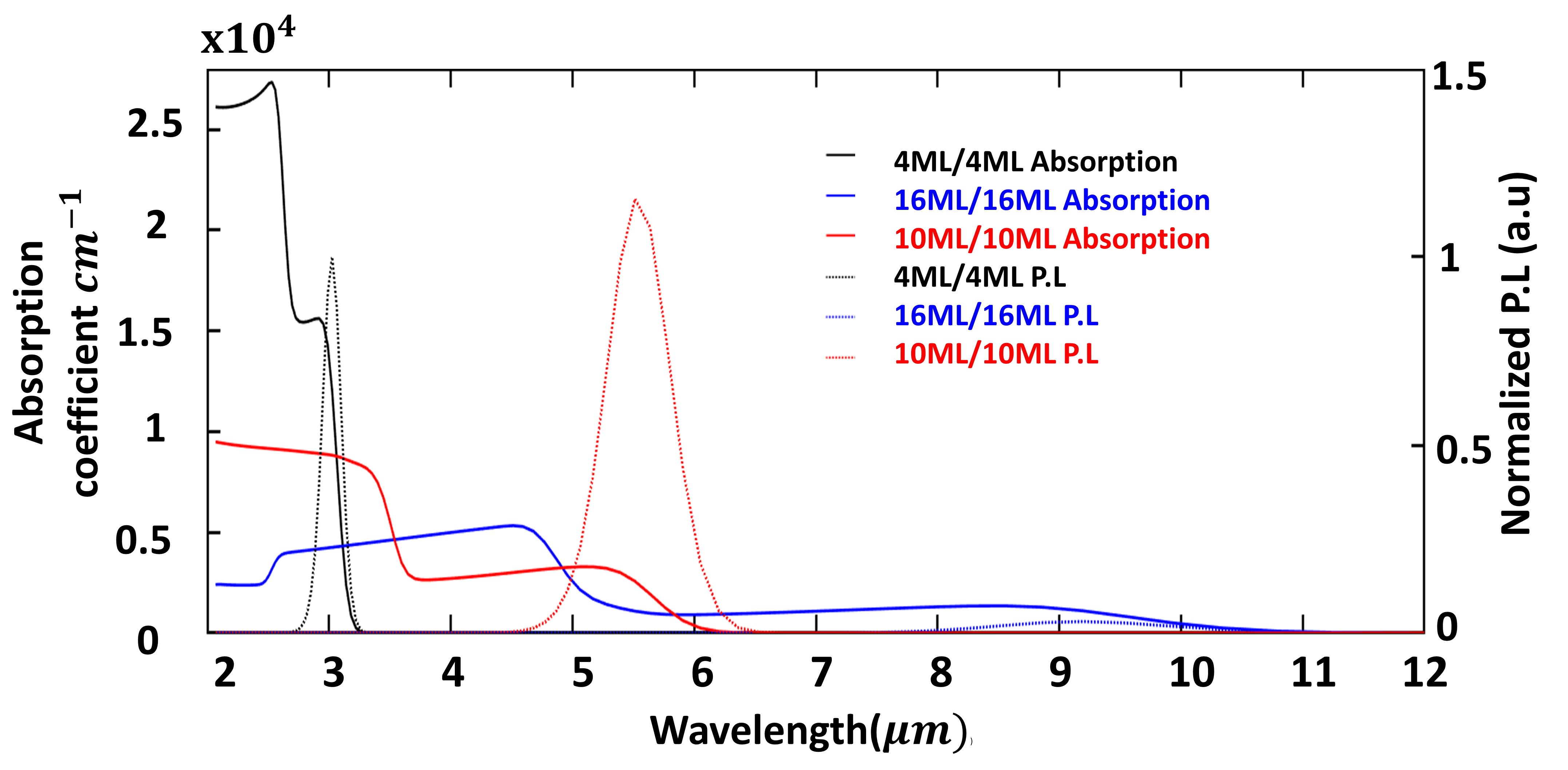}}
%{\includegraphics[height=0.232\textwidth,width=0.46\textwidth]{PL_8_16_A.png}}
	
	\caption{Absorption coefficient and photoluminescence(P.L) calculated by the 8 band $\bf k.p$ and Fermi's Golden for SWIR(4ML/4ML), MWIR(10ML/10ML), and LWIR(16ML/16ML) detectors. The absorption coefficient and Photoluminscene for three T2SLs at T=77K, the absorbers are p-type doped. As we go from LWIR to SWIR, the cutoff wavelength increases as the bandgap decreases with the increase in the widths of InAs and GaSb. The PL peak intensity is lowest for the LWIR, due to higher confinement of the carriers. The P.L plots are normalized to the maximum P.L value of 4ML/4ML T2SL.}
		\label{fig7}
\end{figure}
 \indent Next, we calculate the bandgap, intrinsic carrier concentration ($n_i$), oscillator strength (O.S), SRH, and radiative recombination coefficients for different T2SL compositions while varying the thickness of both InAs and GaSb from 4 ML to 12 ML and 6ML to 22ML, respectively. The bandgap ($E_{g}$) 2D color plot is plotted in Fig. \ref{fig4} (a), and we notice that the bandgap decreases with increasing InAs width (at constant GaSb width) and changes proportionally with GaSb width (at a constant InAs width). Figure. \ref{fig4} (b) shows the variation of $n_i$ with the InAs and GaSb widths. A rapid rise in $n_i$ is seen with the increment in the InAs width as the $E_{g}$ reduces and the effective masses of the holes increase\cite{mukherjee2021carrier,singh2022comprehensive,delmas2019comprehensive}.  
For a constant InAs width while increasing the GaSb width, $n_{i}$ decreases as $E_{g}$ increases (conduction band split reduces, $CB_{min}$ shifts upwards)\cite{singh2022comprehensive}.\\
\indent
In Fig.\ref{fig4} (c), we calculate the oscillator strength (O.S) of the first conduction and heavy hole ($C_{1}$-$HH_{1}$) transition at the Brillouin zone center by using $\int g_{1,n}g_{2,m}dz$, which is the product of the dominant envelope wavefunction for the conduction ($g_{1,n}$) and heavy hole band ($g_{2,m}$). The O.S decreases with the increased width of the InAs and GaSb as the confinement of the carriers in their respective wells increases; therefore, the thicker superlattices have less absorption than the thinner ones. Figures. \ref{fig5} shows the minority electron(p-type absorber) SRH recombination carrier lifetime, calculated by using $\tau_{n0}$ and $\tau_{p0}$, as a function of the thermal velocities of the carriers, which in turn relies on the effective masses ($vth=\sqrt{8k_{B}T/(\pi m_{(e/h)}m_{0})}$)\cite{aytac2016evidence,klein2014defect,aytac2015temperature}. The capture time constants ($\tau_{p0}$ and $\tau_{n0}$) are calculated by equation \ref{tau}. We consider a trap energy level of 10$meV$ below the intrinsic level and the product of cross-sectional area and trap density $\sigma Nt$ equal to $2.4 cm^{-1}$ \cite{taghipour2019temperature} to calculate the SRH lifetime \cite{connelly2010direct}. From Fig. \ref{fig5}, we notice the increment in the carrier lifetime ($\tau_{mc}$) as the InAs(GaSb) width is increased while keeping the GaSb(InAs) constant. \\
%Here, the effective mass of the hole increases with the higher InAs width, causing the thermal velocity of holes to drop and hence the $\tau_{p0}$ (inversely proportional to thermal velocity) to increase. The electron capture time constant ($\tau_{n0}$)  increases with increasing GaSb width for any constant InAs width, because the electrons become more confined in the InAs well, which leads to higher effective mass and thus a lower thermal velocity, as shown in Fig.\ref{fig5} (b).\\
\begin{figure}[!htbp]
\centering
{\includegraphics[height=0.23\textwidth,width=0.45\textwidth]{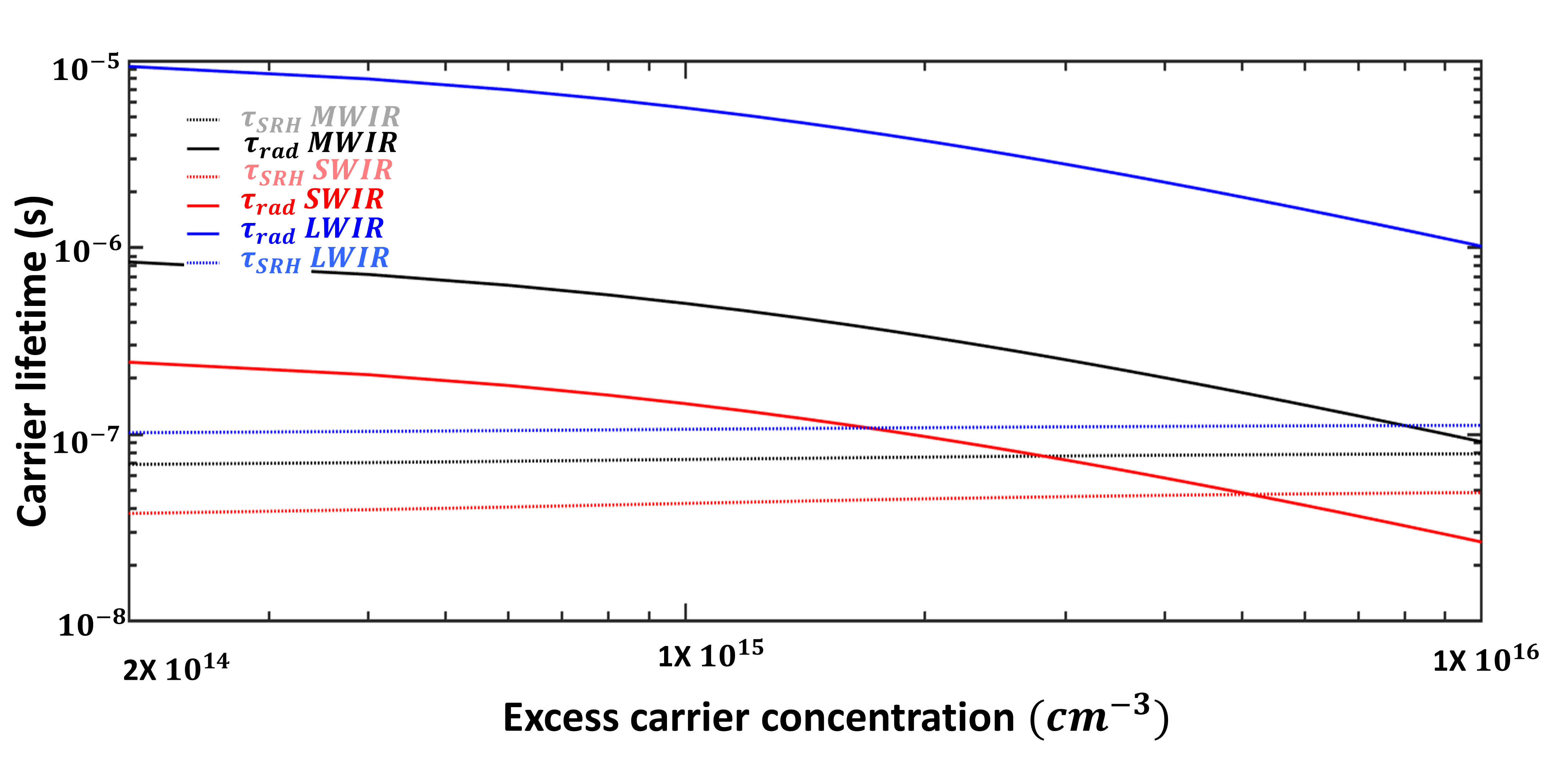}}
   \caption{SRH and radiative carrier lifetime plot calculated by the 8 band $\bf k.p$ and Fermi's Golden rule for SWIR, MWIR, and LWIR T2SLs absorbers with various excess carrier concentrations where all the absorbers are p-type($1x10^{15}cm^{-3}$). The SRH and radiative lifetime are higher for the LWIR absorber as with the increment in the widths of InAS and GaSb, carriers are more localized in their respective layers and their overlap is reducing, therefore, their mini bandwidth is reducing and hence they have lower velocities and higher confinement. At low-level injection, SRH is the dominant phenomenon for all wavelengths.}
		\label{fig8}
\end{figure}
\indent After investigating the SRH recombination carrier lifetimes, which dominate at lower temperatures, we now apply Fermi's golden rule to calculate the absorption coefficient. The required optical matrix moment elements are calculated by the 8-band $\bf k.p$ theory as described in section II.B. We confirm our simulated absorption coefficients with experimental values as shown in Fig. \ref{fig6}, where the
absorption coefficients for the 8.5ML/13.6ML, 14.4ML/7.2ML, and 16ML/7ML T2SL wavelengths, which are approximately $4.7\mu m$, $11\mu m$  and $12.5\mu m$ cut-off wavelength, respectively, are calculated at T=77K.  The results are consistent with the experimentally demonstrated value.\\
\indent After validating our absorption model, we calculate the absorption coefficients for the 4ML/4ML, 10ML/10ML, and 16ML/16ML T2SL with a doping concentration of $10^{15} cm^{-3}$ (p-type doped) to study the variations in the optical properties of T2SL concerning the InAs and GaSb compositions.
The p-type doping is taken into account in the absorber because the minority carriers, which are electrons in p-type, have a longer diffusion length; therefore, they contribute more to photocurrent, and thus a higher quantum efficiency can be achieved. While moving from LWIR to SWIR, we observe a decrease in cut-off wavelength, as shown in Fig.\ref{fig7}. Also by increasing the width of InAs and GaSb, the confinement increases, due to which the carrier overlap at the interface decreases, which reduces the probability of their transition from a filled valence band to an empty conduction band and we notice lower absorption coefficient values in Fig.\ref{fig7} for higher wavelength region.\\
\indent Since the photoluminescence (P.L) spectra correspond to the spontaneous emission rate at lower doping,  we then calculate the P.L spectra of T2SL at low injection levels, where we keep the excess carrier concentration as $10^{14}cm^{ -3} $, and the acceptor's concentration ($N_{A}$) is $10^{15} cm^{ -3}$ and plot the P.L spectra in Fig.\ref{fig7}.
The intensity of the P.L peak decreases as we move from SWIR to LWIR region, which is because the downward transition, i.e., from the filled conduction band to the empty valence band, is smaller when the joint density of states (responsible for absorption also) is less as the spatial overlap of the carriers is less due to higher localization in their respective wells, and therefore, the P.L intensity reduces. The P.L intensity and absorption are also dependent on the quasi-Fermi levels, the P.L intensity increases with acceptor doping, and the absorption decreases with acceptor doping. The values of $f_{c}$ and $f_{v}$ as taken in equation \ref{alpha} are calculated by assuming the low injection by equations \ref{fc} and \ref{fv}.
%We also see the increment decrease in the full width at half maximum (FWHM) as the confinement increases which increases the scattering carrier lifetime, and therefore, lowers FWHM.
Furthermore, using equation \ref{rad}, we estimate the radiative recombination coefficients ($B_{rad}$) for these configurations. The calculated $B_{rad}$ values are $3.4x10^{-9} cm^{3}s^{-1}$, $9.9x10^{-10}cm^{3}s^{-1} $, $8.94x10^{-11}cm^{3}s^{-1}$ for SWIR, MWIR, and LWIR T2SLs, respectively. As $B_{rad}$ lowers, the carrier's lifetime increases; hence, the 16ML/16ML(LWIR) has the longest radiative recombination lifetime.\\
\indent Next we plot the SRH and radiative carrier lifetime for these superlattice in Fig.\ref{fig8} where the absorbers are p-type doped and the excess carrier concentration is varied from $2x10^{14} cm^{-3}$ to $1x10^{16} cm^{-3}$. At low injection level the SRH is dominating phenomena.
%We also see the increment decrease in the full width at half maximum (FWHM) as the confinement increases which increases the scattering carrier lifetime, and therefore, lowers FWHM. Furthermore, using equation \ref{rad}, we estimated the radiative recombination coefficients (B) for these configurations and found that B decreases as the widths of InAs and GaSb increase. As the B lowers, the carrier's lifetime increases, hence, the 16ML/16ML has the longest radiative recombination lifetime.
\begin{figure*}[!htbp]
	\centering
 %\subfigure[]
		{\includegraphics[height=0.25\textwidth,width=1\textwidth]{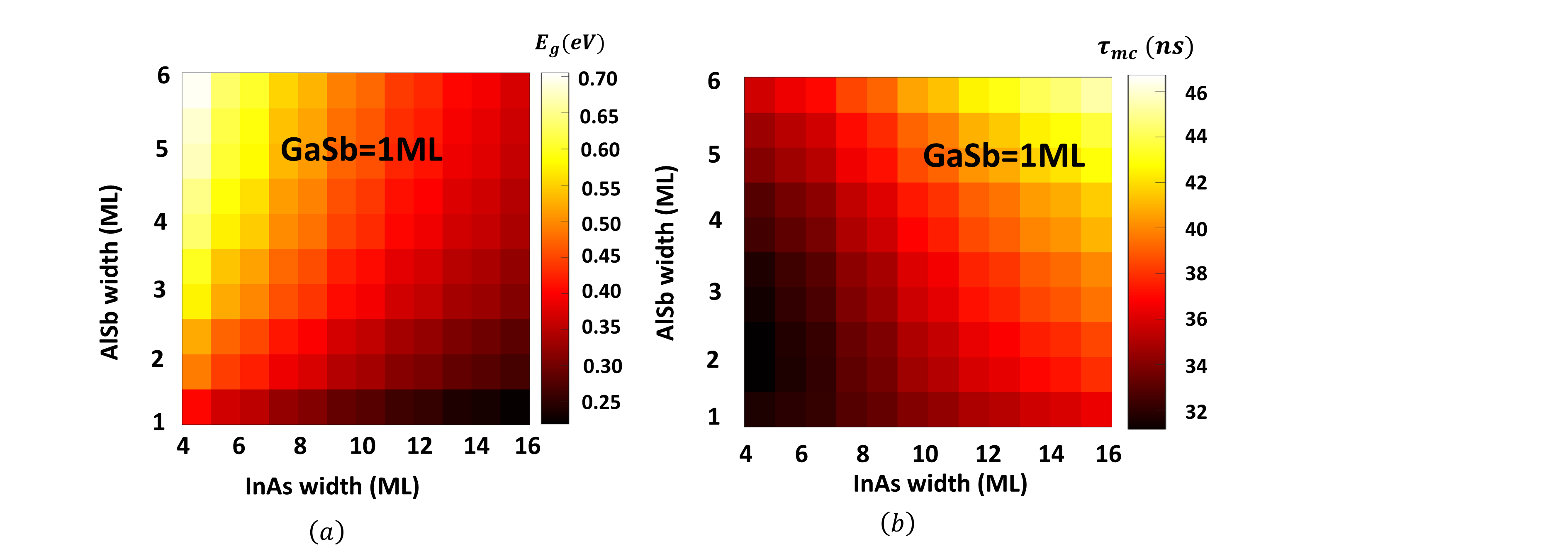}}
	
	\caption{The 2D color plots for the bandgap and the SRH minority carrier lifetime for MSL calculated by the 8 band $\bf k.p$ perturbation theory. (a) Bandgap of MSL when InAs width varies from 4ML to 16ML and AlSb from 1ML to 6ML while keeping the GaSb width constant (1ML). (b) The SRH lifetime is plotted with the InAs width varying from 4ML to 16ML and AlSb width from 1ML to 6ML, keeping GaSb at 1ML. The bandgap increases with the increase in AlSb thickness (InAs constant). Keeping constant InAs width with increase in the AlSb width, there is an increase in the carrier lifetime.}
		\label{fig9}
\end{figure*}
\begin{figure}[!htbp]
\centering
 {\includegraphics[height=0.25\textwidth,width=0.45\textwidth]{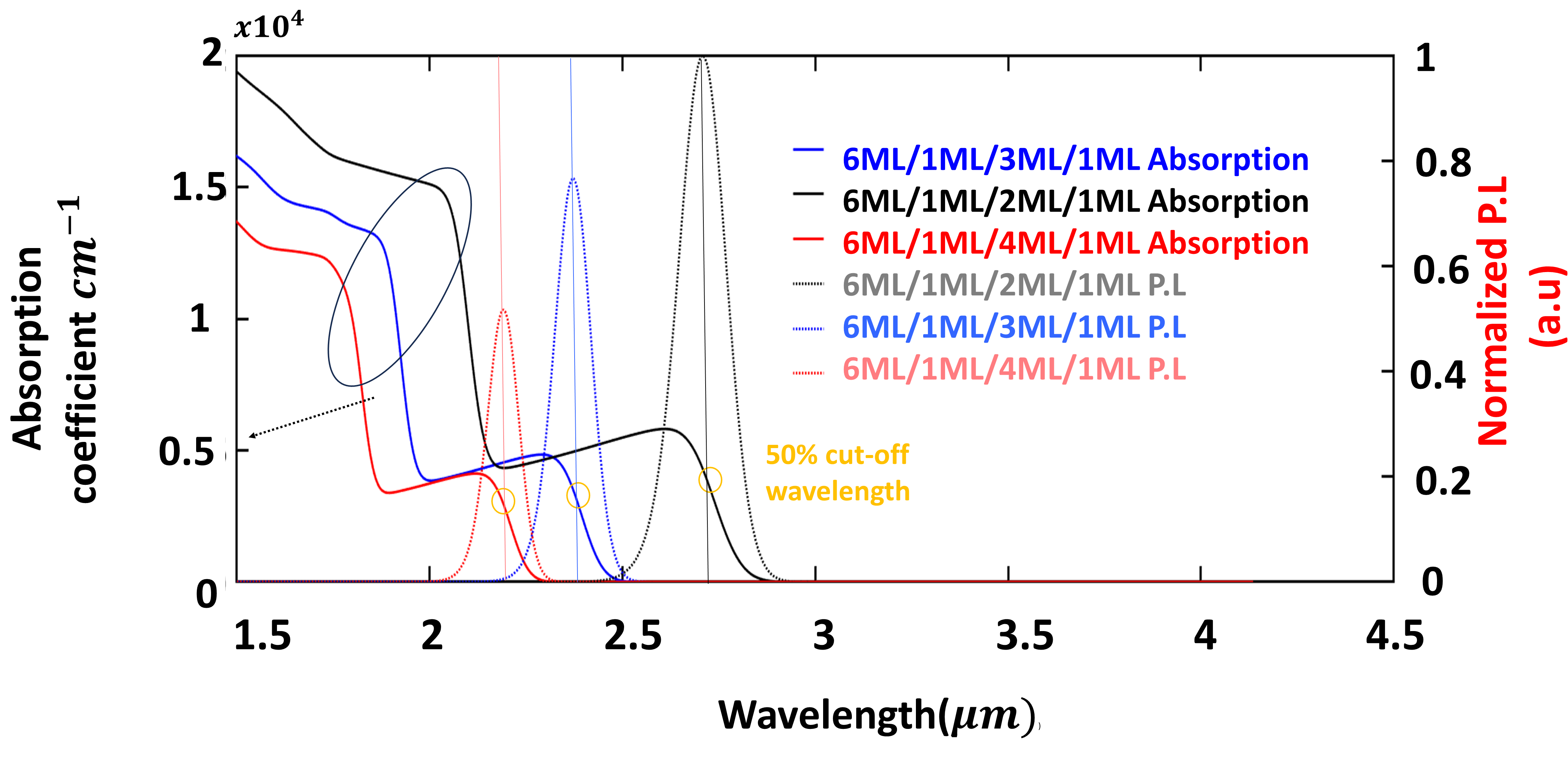}}
   
	\caption{The theoretically calculated TE absorption coefficient and P.L spectra for the 6ML/1ML/xML/1ML MSL where the x=2,3,4 ML. It has been noticed that there is a blue shift in the P.L peak energy with the increase in the AlSb thickness and also the PL peak intensity reduces.}
		\label{fig10}
\end{figure}
\subsubsection{\textbf{InAs/GaSb/AlSb/GaSb M superlattice absorber}}
Since the addition of high-gap AlSb at the GaSb center in the InAs/GaSb/AlSb/GaSb MSL shifts the center of hole wavefunction near the InAs/GaSb interface and thus increases the absorption \cite{lang2013electronic}. We want to study the MSL to gain insight into this fact.
The electronic band properties of MSL, likewise the effective masses, the conduction and valence band edge tunability (with the InAs, GaSb, and AlSb), and density-of-states, we have reported in our previous work \cite{singh2022comprehensive}. Here, we take a constant thickness of 1 ML of GaSb and change the InAs and AlSb widths from 4ML to 16ML and 1ML to 6ML, respectively, to see the effect of the AlSb width on the bandgap and the minority carrier lifetime. From Fig. \ref{fig9} (a), we deduce that the $E_{g}$ increases with the increasing width of AlSb because AlSb acts as an additional barrier to the electrons in the consecutive InAs wells, which further reduces the conduction band split and thus leads to an increase in $E_{g}$. By taking the p-type absorber($10^{15}cm^{-3}$) and under low-level injection, we notice that the SRH lifetime increases with the AlSb thickness at a constant InAs thickness as shown in the Fig. \ref{fig9} (b).\\
\indent Next, in Fig. \ref{fig10}, the absorption and P.L spectra are plotted for the 6ML/1ML/xML/1ML MSL, where the x varies from 2ML to 4ML, to notice the effect of AlSb width. The absorption coefficient values at the 50 \% cut-off wavelength decrease with increased AlSb width. Also, the reduction in the 50\% cut-off wavelength with the increase in the AlSb is seen in addition to the reduced P.L intensity. It is noticed that the Full-width half maxima (FWHM) decreases with the increase in the AlSb width. The higher FWHM denotes the higher scattering and, hence, less carrier lifetime. Here, with the increase in the AlSb width, as carriers' effective masses increase\cite{singh2022comprehensive}, they are less scattered in the material; therefore, there is a reduction in the FWHM with the increase in the AlSb width. The radiative recombination rate varies from $9.49x10^{-9} cm^{3}s^{-1}$ to $5.4x10^{-9}cm^{3}s^{-1}$ while increasing the AlSb width from 2ML to 4ML.\\
\indent Since we can achieve the same bandgap with different compositions of T2SL and MSL compositions, we want to investigate the effect of InAs, GaSb, and added AlSb (in the case of MSL) on the optical and dark current properties of T2SL and MSL. Therefore, in the following sections, we compare the two same bandgap T2SL and MSL absorbers for the SWIR and MWIR infrared photodetection.
%It has been noticed that the FWHM decreases with the increase in the AlSb width. The higher FWHM denotes the higher scattering and hence, less lifetime. Here, with the increase in the AlSb width, as carriers' effective masses increase\cite{singh2022comprehensive}, they are less scattered in the material, therefore is a reduction in the FWHM.  
%HH-LH band mixing leads to the carrier scattering in the minibands, higher FWHM.
\begin{figure*}[!htbp]
	\centering
			\subfigure[]{\includegraphics[height=0.25\textwidth,width=0.45\textwidth]{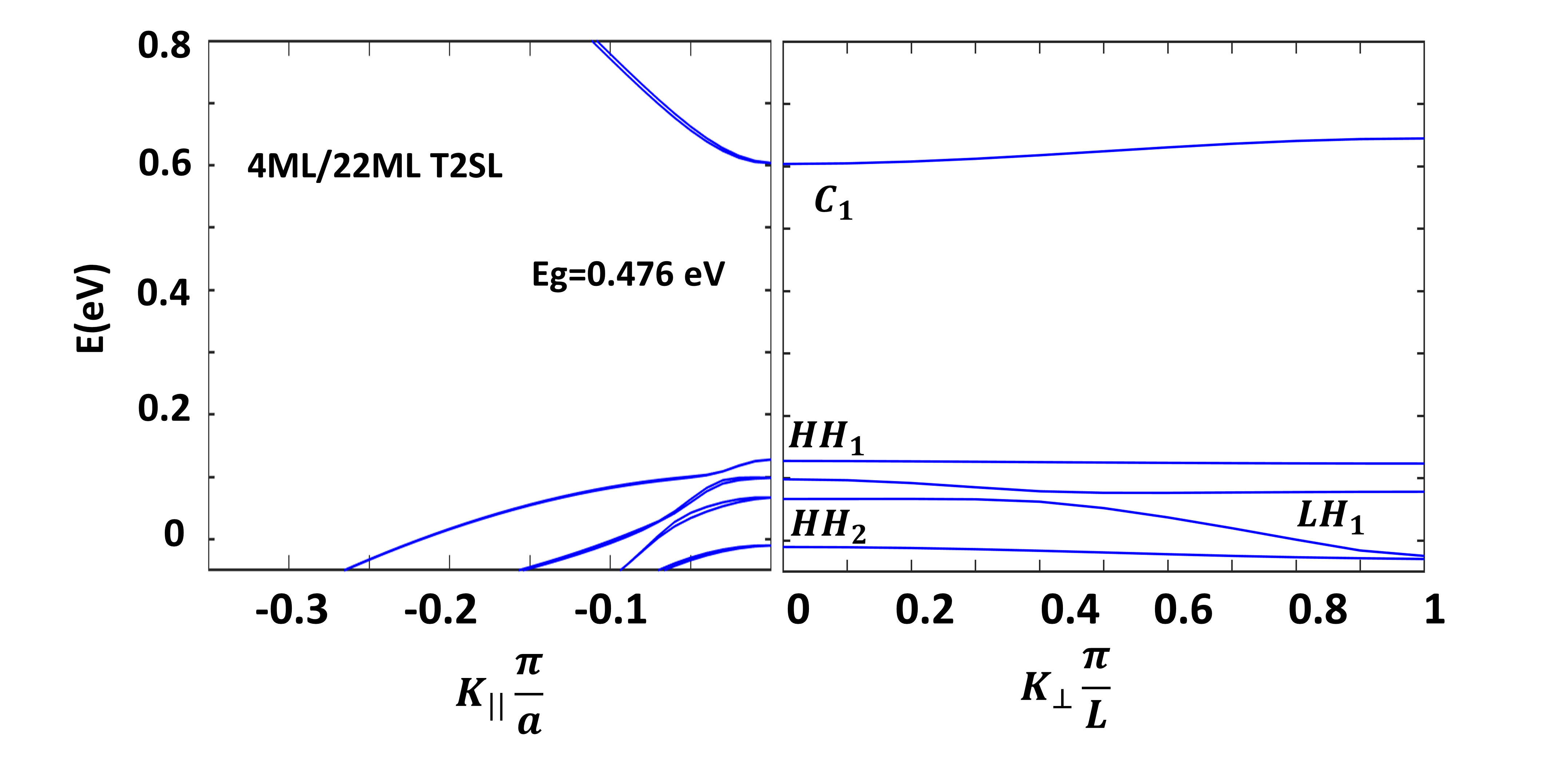}}
   \subfigure[]{\includegraphics[height=0.25\textwidth,width=0.45\textwidth]{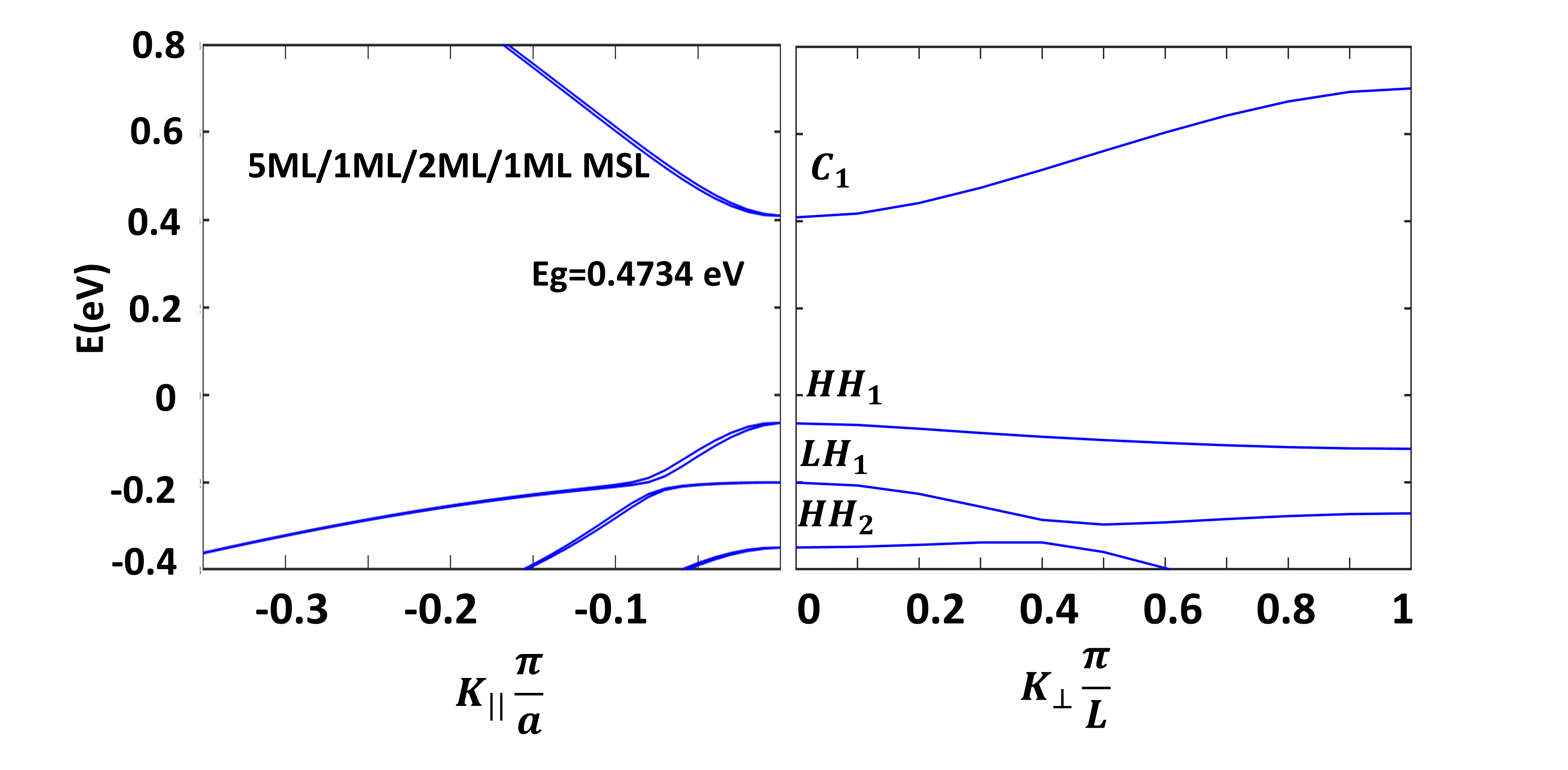}}		
		\caption{Energy dispersion plot calculated by the 8 band $\bf k.p$ theory for the T2SL and MSL at T=77K. (a) E-k for both in the parallel and out-of-the plane for 4ML/22ML T2SL(b)  E-k for both in the parallel and out-of-the plane for 5ML/1ML/2ML/1ML MSL. The mini bandwidth for the electrons in the T2SL and MSL are approximately 135$meV$ and 297$meV$, respectively. Therefore, the electrons are highly confined in the InAs quantum well in the T2SL.}
		\label{fig11}
\end{figure*}

\begin{figure*}[!htbp]
	\centering
 \subfigure[]
		{\includegraphics[height=0.25\textwidth,width=0.45\textwidth]{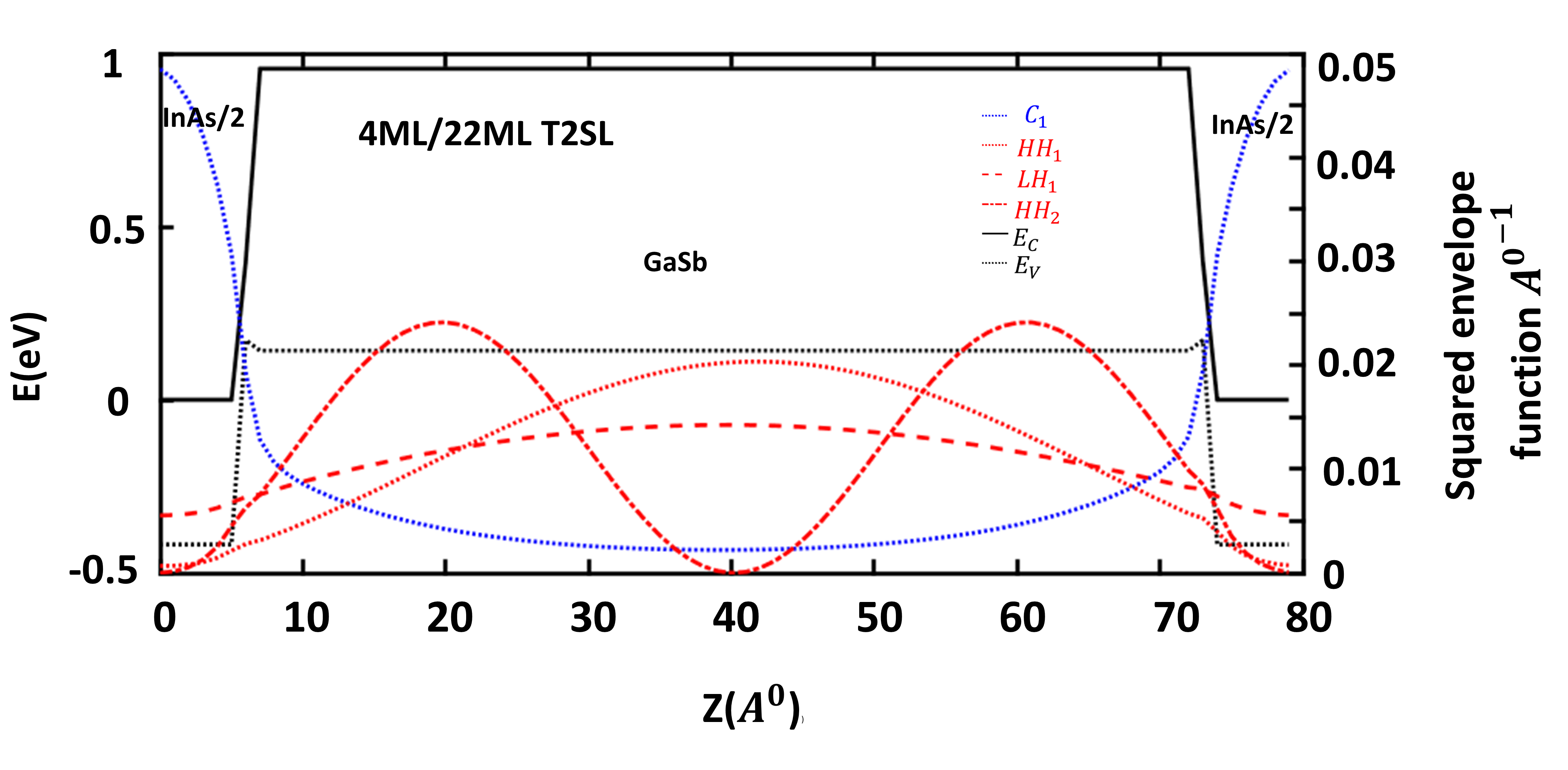}}
   \subfigure[]
		{\includegraphics[height=0.25\textwidth,width=0.45\textwidth]{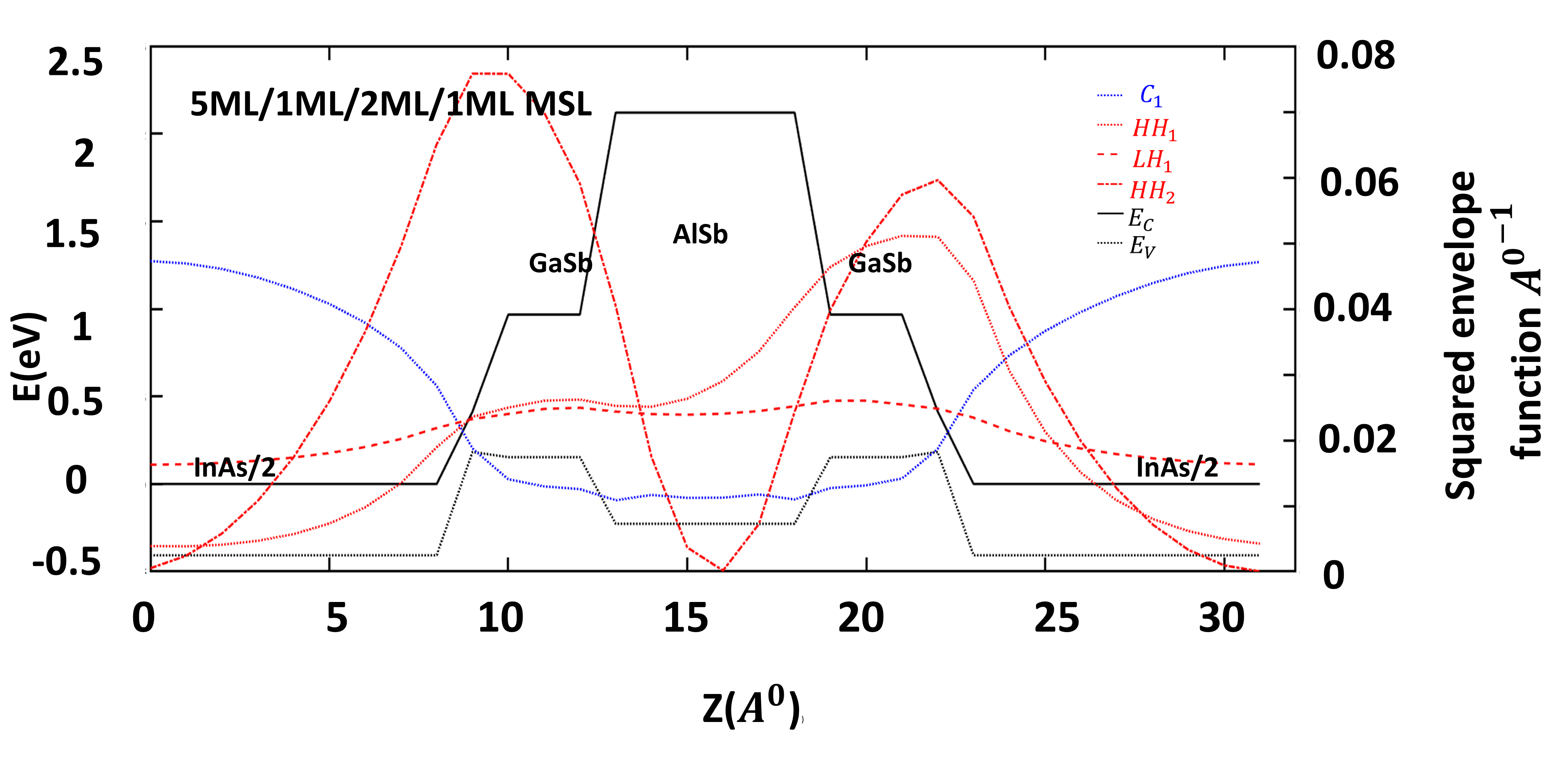}}
		\caption{Squared envelope wavefunction and the energy band alignment of the two same bandgap SWIR superlattices at zero applied voltage. We take the InSb(GaAs) layer at the interface of InAs/GaSb(GaSb/InAs) in the T2SL and  InSb at both of the interfaces in MSL. (a) 4ML/22ML T2SL (b) 5ML/1ML/2ML/1ML MSL. }
		\label{fig12}
\end{figure*}
\begin{figure*}[!htbp]
	\centering
		{\includegraphics[height=0.25\textwidth,width=1\textwidth]{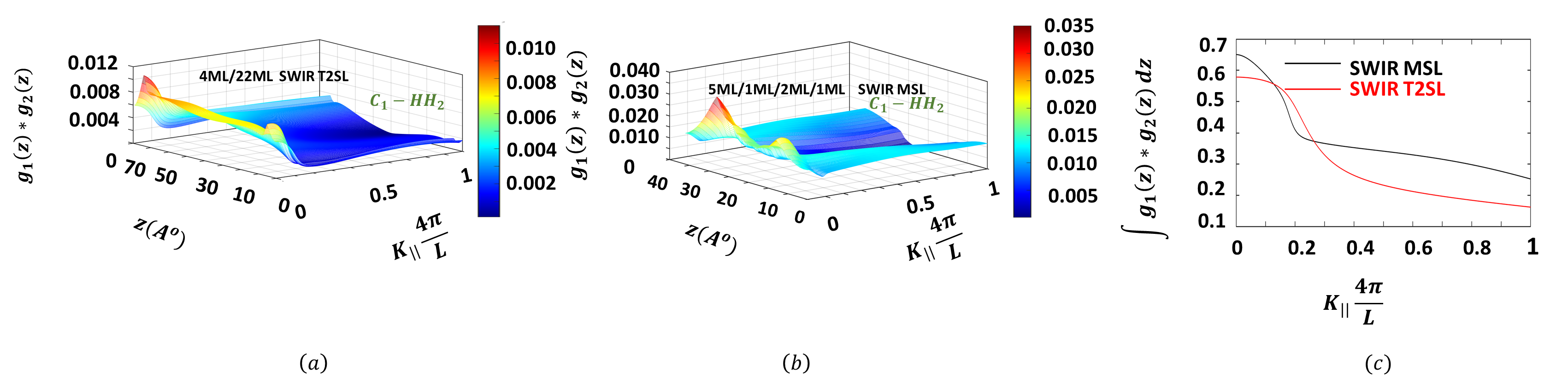}}
		\caption{ Spatial product of envelope wavefunction of first conduction and first heavy hole bands w.r.t to the growth direction and in-plane wave vector and oscillator strength at zero applied bias (a) 4ML/22ML (b) 5ML/1ML/2ML/1ML (c) Oscillator strength for the (a) and (b). As the carriers are more delocalized in the MSL, their finding probabilities at the interface are higher which led to the higher overlap at the InAs/GaSb interface in the MSL.}
		\label{fig13}
\end{figure*}
\begin{table}
    [!htbp]
\caption{Properties of superlattices calculated for the comparison of two same bandgap short wavelength superlattices by the 8 band $\bf k.p$ perturbation at $T=77K$.} 
\label{table2}
\begin{center}      
\begin{tabular}{|l|l|l|l|l|}
\hline
\rule[-1ex]{0pt}{3.5ex} 	Parameters & T2SL& MSL\\
\hline
\rule[-1ex]{0pt}{3.5ex} Bandgap (eV) & 0.476 & 0.4738 \\
\hline
\rule[-1ex]{0pt}{3.5ex}Overall Electron effective mass($m_{0}$) & 0.0495 & 0.0459\\
\hline
\rule[-1ex]{0pt}{3.5ex}Overall Hole effective mass($m_{0}$) & 0.0870& 0.0604\\
\hline
\rule[-1ex]{0pt}{3.5ex} Intrinsic carrier concentration $n_{i} (m^{-3})$   & $2.26x10^{6}$ & $1.98x10^{6}$\\  
\hline
\rule[-1ex]{0pt}{3.5ex} Radiative recombination coefficient ($B_{rad}$) & $2.5x10^{-11}$ & $3.03x10^{-12}$ \\  

\hline
\end{tabular}
\end{center}
\end{table}
\begin{figure}[!htbp]	
\centering
{\includegraphics[height=0.25\textwidth,width=0.55\textwidth]{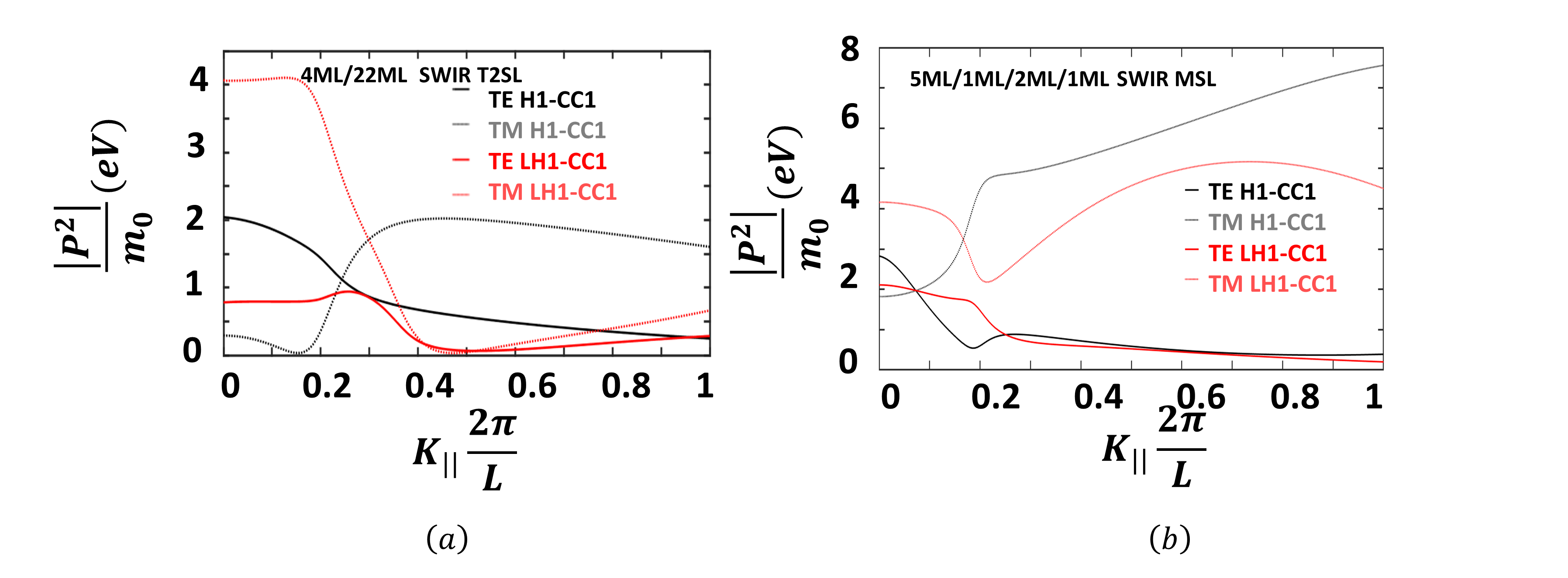}}\label{b1}
	\quad
			\caption{Theoretical TE and TM optical moment matrix element calculated for the 4ML/22ML T2SL and 5ML/1ML/2ML/1ML MSL which has the same bandgap of around 0.476 $eV$. (a) TE and TM transition for 4ML/22ML T2SL (b) TE and TM transitions 5ML/1ML/2ML/1ML MSL. The $C_{1}-HH_{1}$ z polarization (TM) is lower than the x polarization (TE) near the Brillouin zone center for both the superlattices.}
		\label{fig14}
\end{figure}
\begin{figure}[!htbp]	
	\quad
 {\includegraphics[height=0.23\textwidth,width=0.45\textwidth]{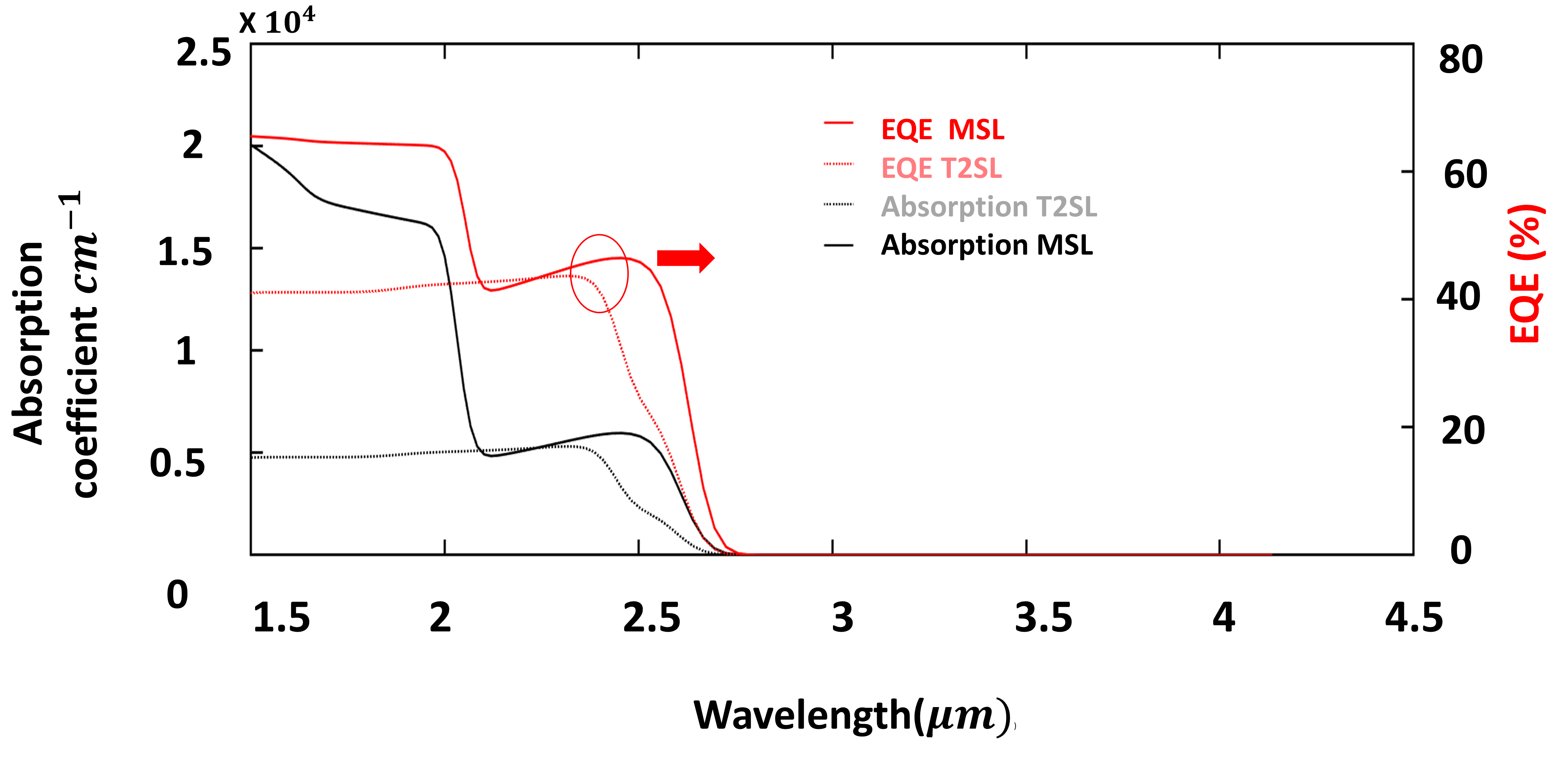}}

			\caption{Theoretical absorption spectra and external quantum efficiency (EQE) calculated by using Fermi's golden rule for short wavelength 4ML/22ML T2SL and 5ML/1ML/2ML/1ML MSL.}
		\label{fig15}
\end{figure}
\begin{figure}[!htbp]
	\centering
{\includegraphics[height=0.22\textwidth,width=0.4\textwidth]{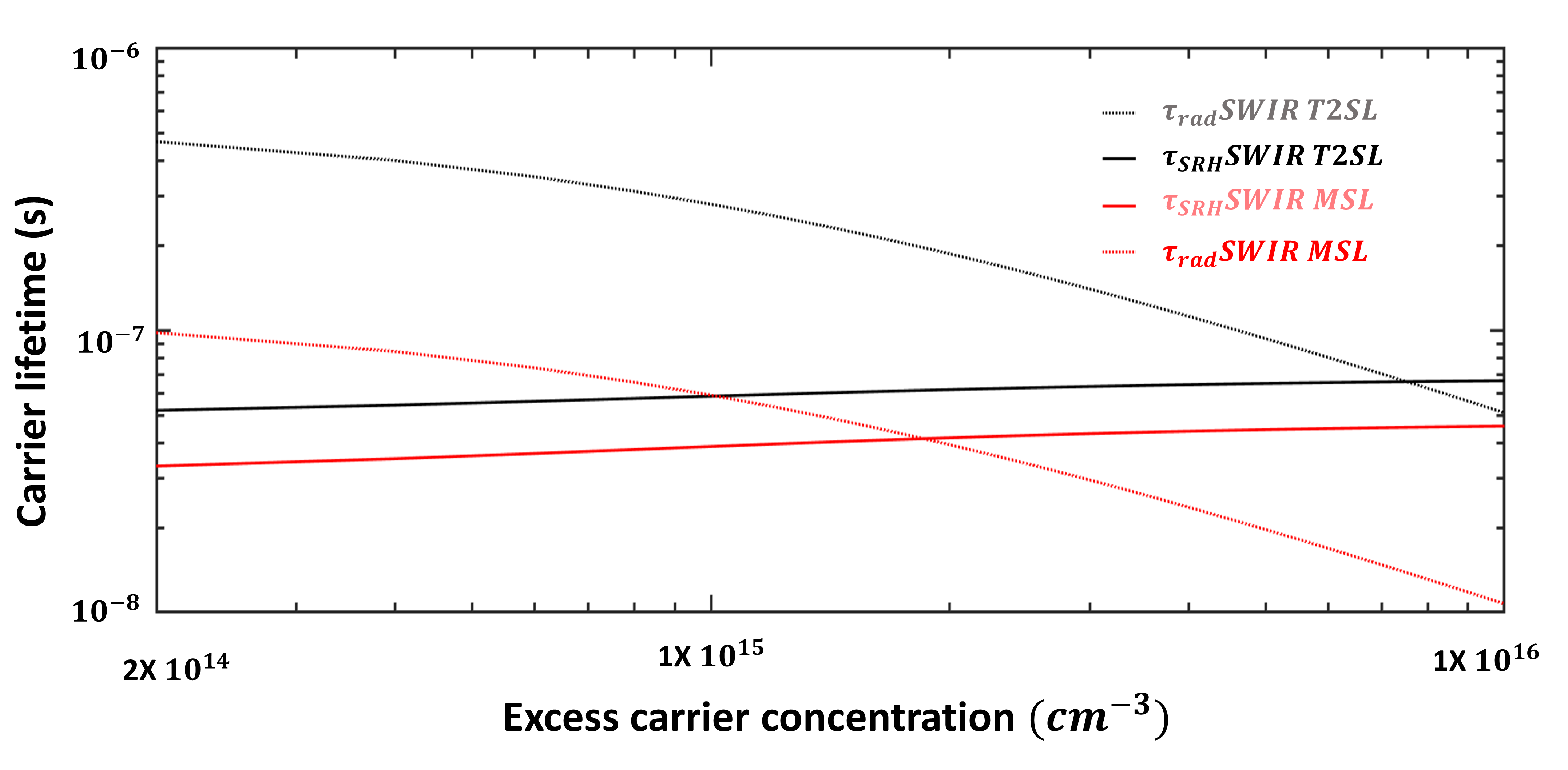}}
  
		\caption{The radiative and SRH lifetimes calculated by the 8 band $\bf k.p$ method for the short wavelength T2SL and MSL when the excess carrier concentration is varied from $2x10^{14}-1x10^{16}cm^{-3}$.}
		\label{fig16}
\end{figure}
\subsubsection{\textbf{Comparison for SWIR Absorber}}
To examine the MSL as an absorber for SWIR applications, we design 5ML/1ML/2ML/1ML MSL and 4ML/22ML T2SL, which have the same bandgap of around 0.476$eV$. The band structures of 5ML/1ML/2ML/1ML MSL and 4ML/22ML T2SL are shown in Fig.\ref{fig11} (a) and (b), both in-plane and out-of-plane directions.  For T2SL and MSL, the observed miniband width is approximately 135$meV$ and 297$meV$, respectively. To capture the overlap of the electron and hole wavefunctions at the interface, we consider the InAs/GaSb structure as (InAs/2)/GaSb/(InAs/2); thus 4ML/22ML is taken as 2ML/22ML/2ML. Similarly, the 5ML/1ML/2ML/1ML MSL is taken as 2.5ML/1ML/2ML/1ML/2.5ML.
%We have taken the alternate InSb and GaAs with 1.5\AA thickness at the InAs/GaSb interface for T2SL, whereas for MSL, InSb with 1.5\AA is taken at each InAs/GaSb interface.
We plot the envelope wave functions corresponding to the first electron, the first heavy hole, the first light hole, and second heavy-hole at the Brillouin zone center ($k_{t,\perp}=0$) in Fig .\ref{fig12}(a) and (b). These envelope wave functions are calculated from the Hamiltonian as discussed in section II.A. The
carrier type's assignment depends on the envelope functions' associated characteristics at $k_{t}=0$\cite{li2010intrinsic}. From Fig.\ref{fig12}(a) and (b), it can be seen that the hole has a higher probability amplitude in the GaSb well, while the electron has more in the InAs electron well.\\
\indent Also, in MSL, the AlSb has separated the GaSb hole quantum well into two newly generated wells for holes. The hole envelope wavefunction is shifted towards the InAs/GaSb interface\cite{lang2013electronic}, as shown in Fig.\ref{fig12} (b). The heavy-hole probability is higher than the conduction band in MSL , whereas, it is lower for T2SL. Moreover, the light-hole occupation is less probable within the GaSb well, and subsequently, its leakage into the InAs barrier is higher than the heavy hole. Each wavefunction consists of eight envelope wavefunction components, which include one conduction and three valence bands plus their spins \cite{galeriu2005k}. These components are denoted by $g_{1,n}-g_{4,m}$ for the upper Hamiltonian, and $g_{5,n}-g_{8,m}$ corresponds to the lower Hamiltonian \cite{ahmed2016analysis}. The $m$ and $n$ are the notation for the conduction and valence bands, respectively. Now, we plot the spatial product, i.e., $g_{1,n}(z)*g_{2,m}(z)$, in Fig. \ref{fig13} (a) and (b). The $g_{1}$ and $g_{2}$ are the strongest components \cite{ahmed2016analysis} of the envelope wavefunction of the first conduction ($C_{1}$) and heavy hole ($HH_{1}$). Here, we observe a higher spatial overlap near the InAs/GaSb interfaces in MSL than in the T2SL. This spatial overlap decreases as it moves away from the zone center ( $k_{t}\neq 0$).
Next, the oscillator strength is calculated as $\int g_{1,n}*g_{2,m}dz$, depict the transition probability from the $HH_{1}$ to the $C_{1}$ \cite{ahmed2016analysis}.
In Fig.\ref{fig13} (c), we notice a higher oscillator strength for MSL than in T2SL.\\ 
\indent Further, we calculate the TE and TM momentum matrix elements to understand the allowed transitions from the first two valence bands to the first conduction band. We plot the TE and TM matrix elements as shown in Figs.\ref{fig14} (a) and (b). The dominant TE transitions are due to the $C_{1}$-$HH_{1}$ transition \cite{andreev2002spin}. It can be observed that $C_{1}$-$LH_{1}$ is the strongest transition in the TM instead of $C_{1}$-$HH_{1}$.\\
 \indent The TE component decreases in the C1-HH1 and the TM component increases from $k_{t}=0$ to $k_{t}=2\pi/L$ \cite{wang2016enhancement}. Also, the calculated TM $C_{1}$-$LH_{1}$ transitions are larger than  TE transitions for both  T2SL and MSL. This is explainable by the fact that the electron-transitioning from the $HH_{1}$ with the angular momentum J=3/2, provides the filling of one electronic spin state in the conduction band\cite{wang2016enhancement,andreev2002spin,subashiev2004optical}. Whereas, with the increase in the excitation the transitioning from the light-hole (J=1/2) also starts contributing to the absorption and filling of the second spin state in the conduction band\cite{subashiev2004optical}.  The $C_{1}$-$HH_{2}$ contributes the least to the absorption spectra, therefore not shown here\cite{wang2016enhancement}. The calculated electron and hole effective masses,
ni and radiative recombination coefficients, are shown in Table. \ref{table2} for these superlattices.
In Fig. \ref{fig15} , we calculate the absorption coefficients for one period of both the T2SL and MSL and notice that the MSL has a higher absorption, which is related to its higher interface overlap and oscillator strength, as discussed above. In Fig. \ref{fig16} we plot the carrier lifetime mainly SRH and radiative for one period of these superlattices, where we vary the excess carrier concentration from $2x10^{14}-1x10^{16}cm^{-3}$ and kept the absorber doping as $1x10^{15}cm^{-3}$. The obtained carrier lifetime is around $38ns$ and $58ns$, at excess carrier concentration of $1x10^{15}cm^{-3}$ for MSL and T2SL, respectively. Next, we calculate the diffusion limited dark current ratio for the T2SL and MSL by utilizing \eqref{jd} on the basis of calculated $n_{i}$, minority carrier lifetime, for $2\mu m$ thickness of the absorber (255 periods of T2SL and 370 periods of MSL) and we found that the $J_{diff}(MSL)$ is 1.17 times of $J_{diff}(T2SL)$, which is approximately similar. Furthermore, We calculate the quantum efficiency by $\eta=(2/3)(1-e^{-\alpha*L}$)(plotted in Fig.\ref{fig15}), where $\alpha$ is the absorption coefficient and the L is the thickness of the absorber which is taken as $2\mu m$ for both T2SL and MSL. At the wavelength around $2.5 \mu m$, the quantum efficiency for MSL and T2SL are 47\% and 26\%, respectively. Therefore, the MSL shows better properties than the T2SL for SWIR photodetectors.  \\
\FloatBarrier
\subsubsection{\textbf{Comparison for MWIR Absorber}}
\begin{figure*}[!htbp]
	\centering
			\subfigure[]{\includegraphics[height=0.25\textwidth,width=0.45\textwidth]{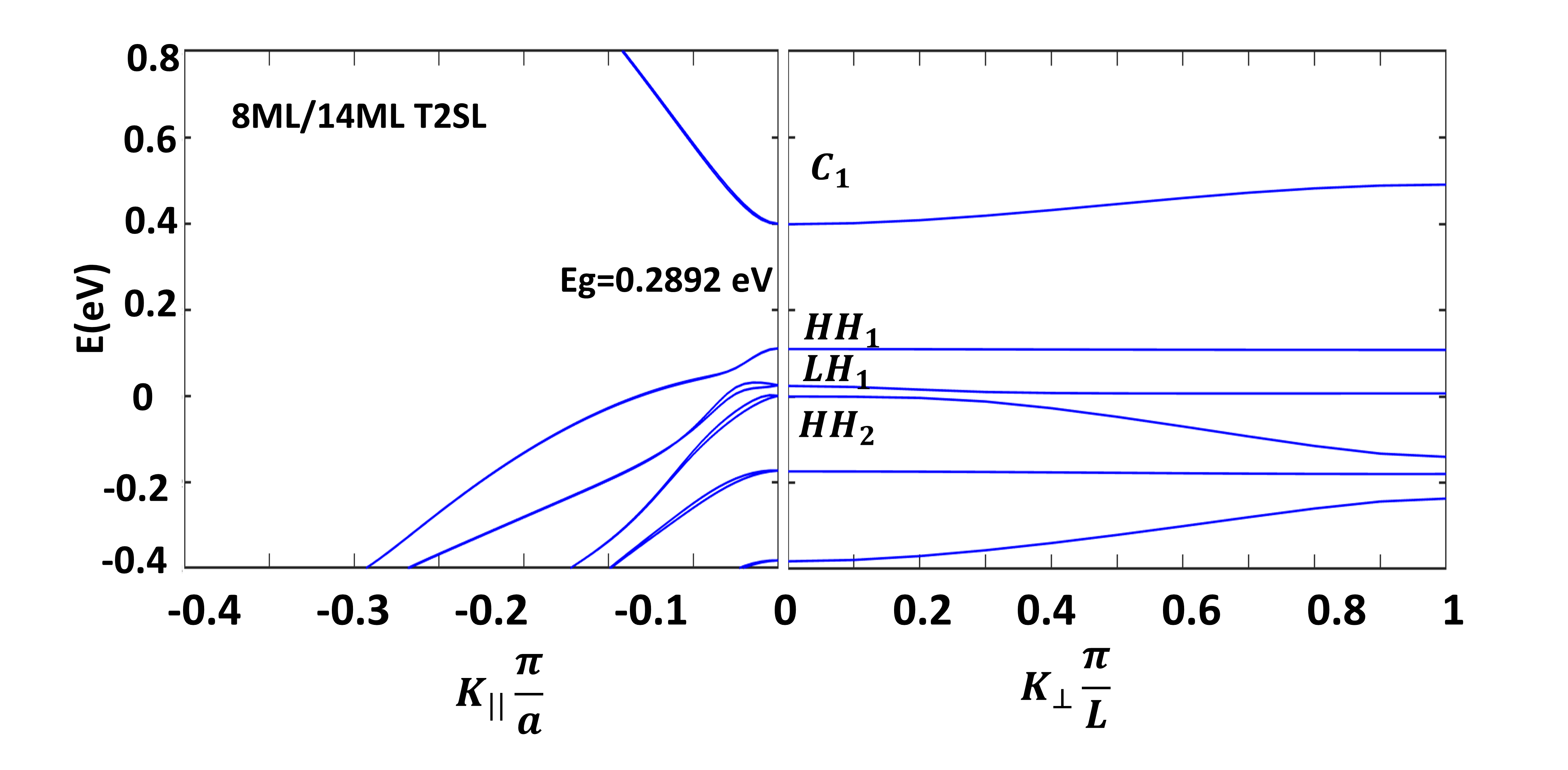}}
   \subfigure[]{\includegraphics[height=0.252\textwidth,width=0.45\textwidth]{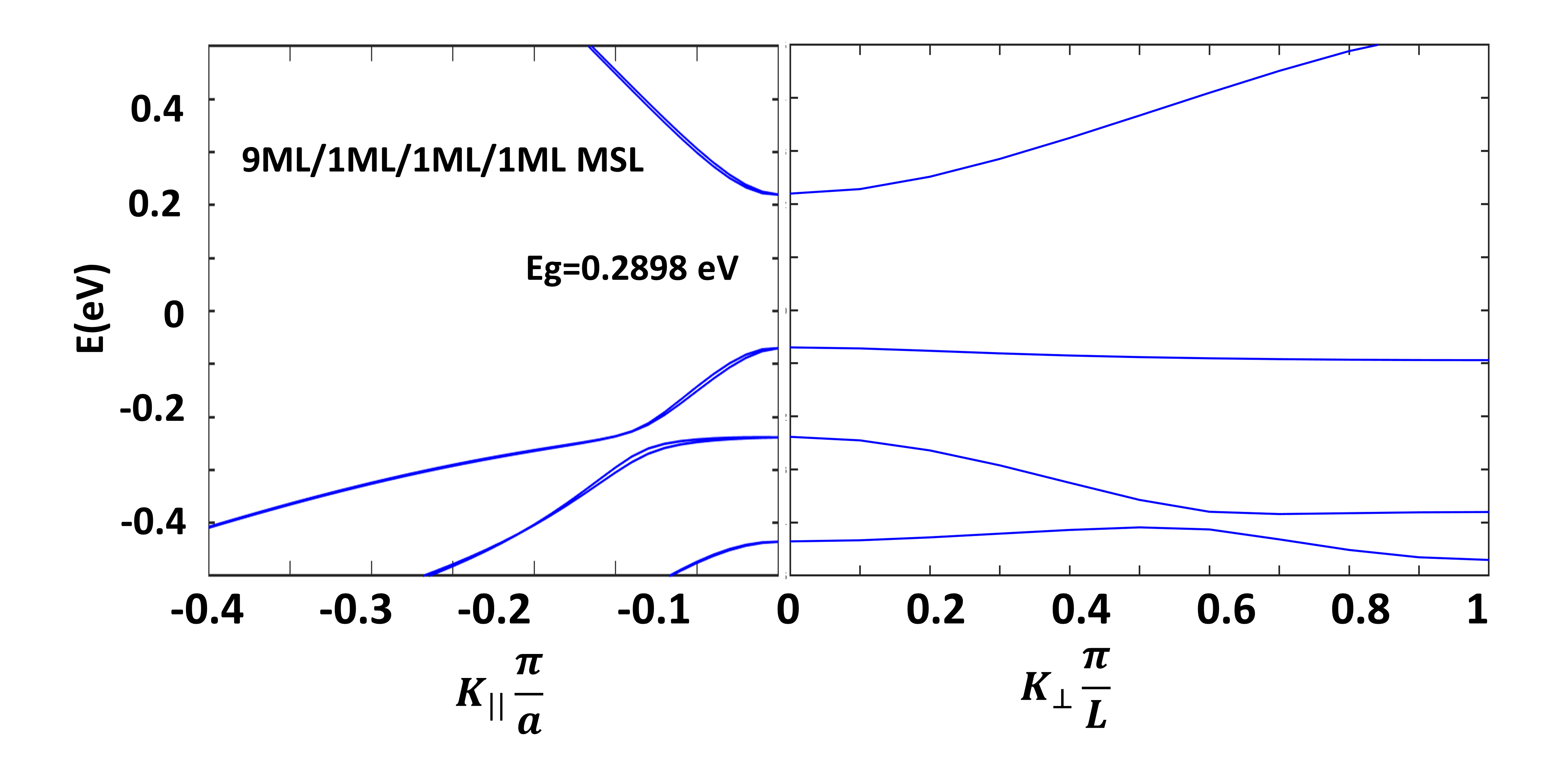}}		
		\caption{Energy dispersion plot calculated by the 8 band $\bf k.p$ theory for the mid wavelength MSL and T2SL at T=77K. (a) E-k for both in parallel and out of the plane for 8ML/14ML T2SL.(b)  E-k for both in parallel and out of the plane for 9ML/1ML/1ML/1ML MSL. The electron mini bandwidth for T2SL and MSL is approximately 96$meV$ and 308$meV$, respectively. Therefore, the electron confinement is higher in T2SL.}
		\label{fig17}
\end{figure*}

 In this section, we compare the MWIR T2SL and MSL absorbers, similar to the SWIR absorber for infrared photodetection. For this, 8ML/14ML T2SL and 9ML/1ML/1ML/1ML MSL configurations are taken, with the equivalent bandgap of 0.289 $eV$, and the corresponding wavelength is around 4.5 $\mu m$. We plot the electronic band structures of these superlattices in Figs. \ref{fig17} (a) and (b). The conduction miniband widths are $96meV$ and $308meV$ for T2SL and MSL, respectively. The holes have negligible mini bandwidth, showing dispersion-less characteristics along the growth
axis \cite{klipstein2021type}. This implies that while heavy holes are extremely localized within the GaSb well, their tunneling transport is less. In contrast, electrons have a wider miniband than heavy holes, and as a result, they contribute to the tunneling current.
\begin{figure*}[!htbp]
	\centering
 \subfigure[]
		{\includegraphics[height=0.25\textwidth,width=0.45\textwidth]{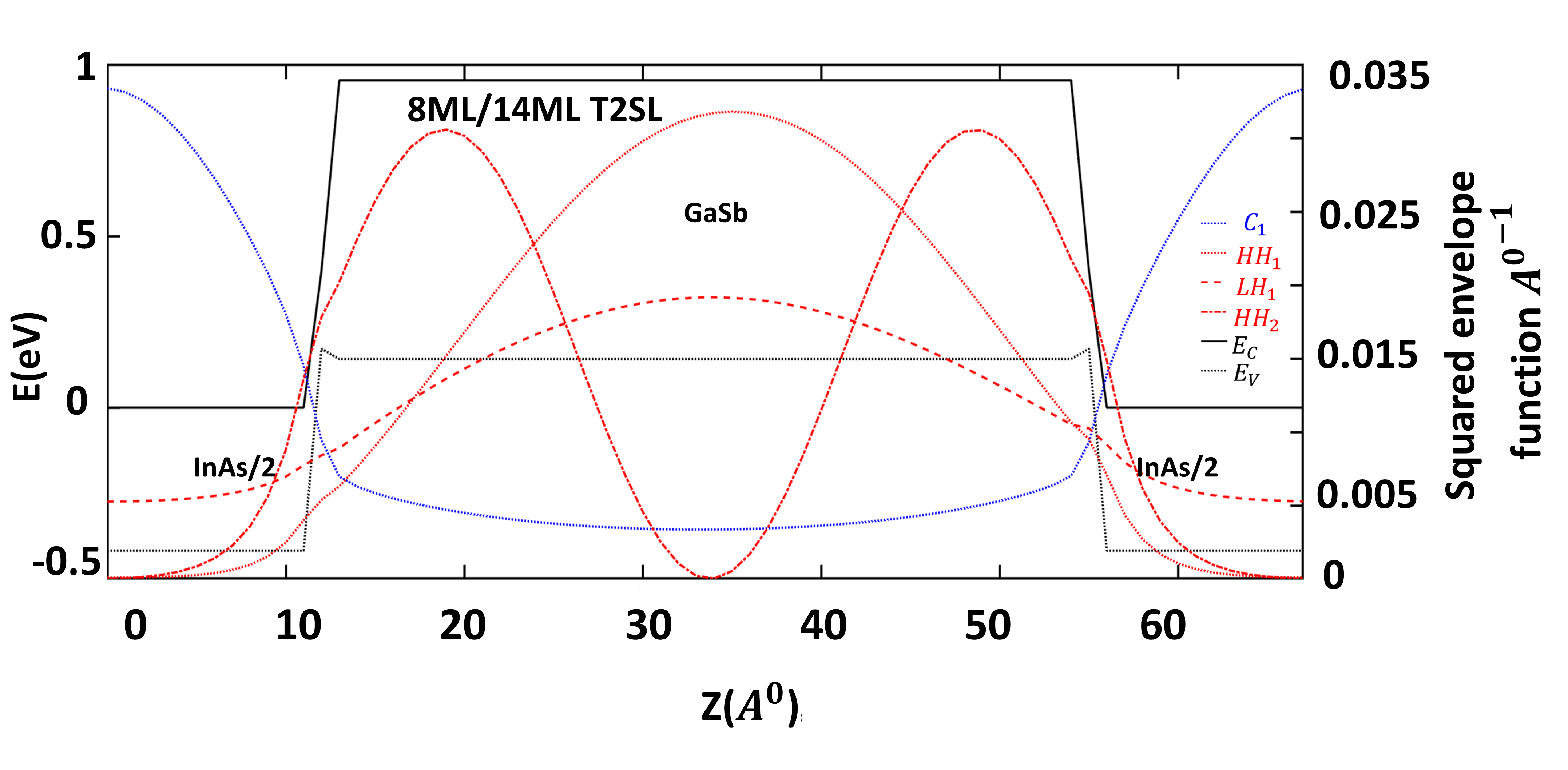}}
   \subfigure[]
		{\includegraphics[height=0.25\textwidth,width=0.45\textwidth]{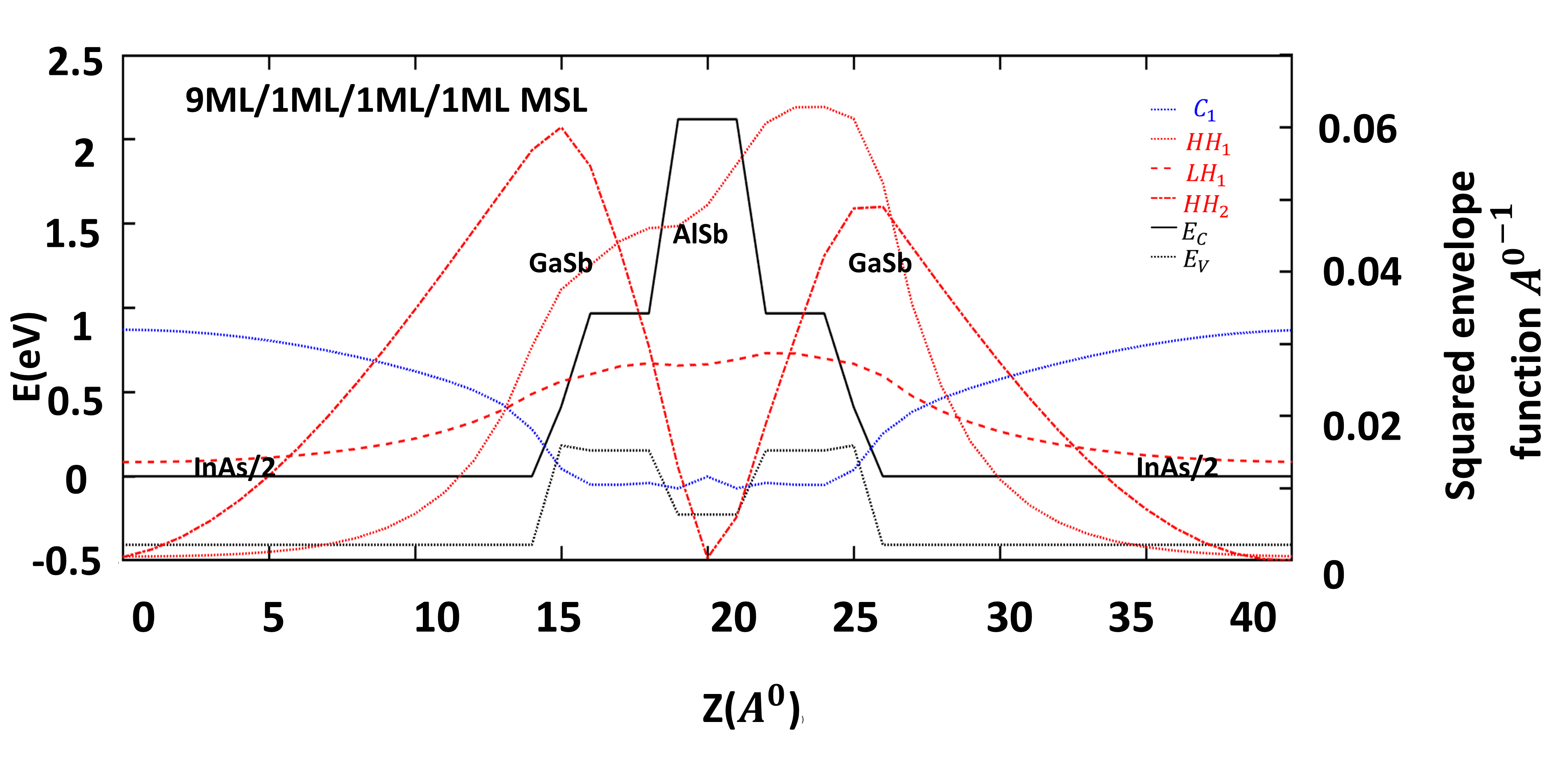}}
		\caption{Squared envelope wavefunction for the conduction and valence subband and the energy band alignment of the two same bandgap superlattices at zero applied voltage w.r.t to the growth axis. (a) T2SL (b) MSL. The squared envelope wavefunctions depict the probability of finding electrons and holes in their respective wells and at the interfaces. }
		\label{fig18}
\end{figure*}
 \begin{figure*}[!htbp]
	\centering
{\includegraphics[height=0.25\textwidth,width=1\textwidth]{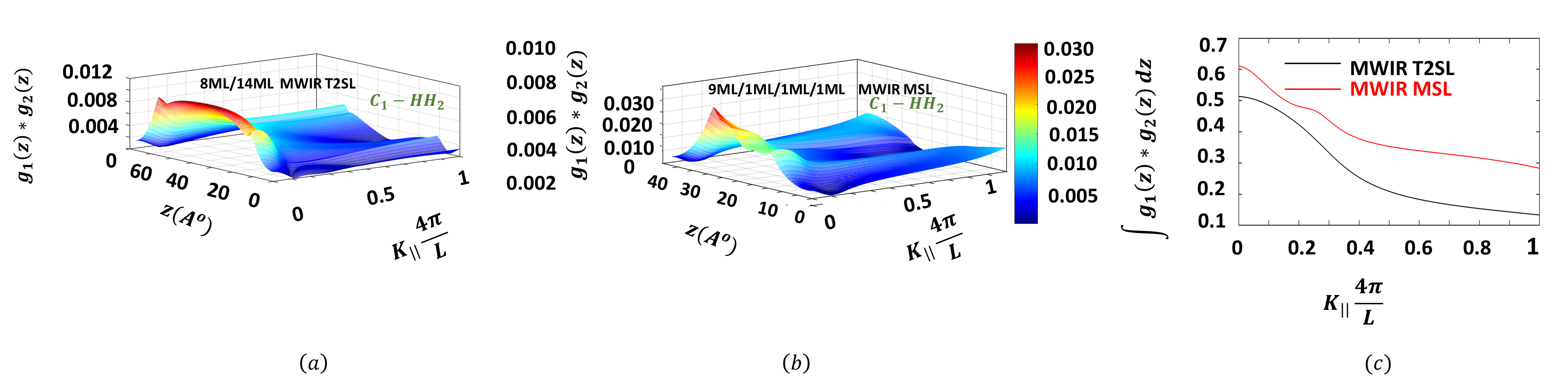}}
		\caption{Product of envelope wavefunction and oscillator strength of first conduction and first heavy hole bands w.r.t to the growth direction (z) and in-plane wave vector at zero applied bias (a) 8ML/14ML T2SL (b) 9ML/1ML/1ML/1ML MSL (c) Oscillator strength of $C_{1}-HH_{1}$ transition for (a) and (b). The AlSb at the center of GaSb in the MSL has pushed the heavy holes more towards the interface therefore, their overlap w.r.t to the growth axis is higher in MSL than in the T2SL. The oscillator strength which is calculated by the integration of the overlaps shown in (a) and (b) at every $K_{||}$ is higher for MSL.}
		\label{fig19}
\end{figure*}
\begin{figure}[!htbp]	
{\includegraphics[height=0.25\textwidth,width=0.5\textwidth]{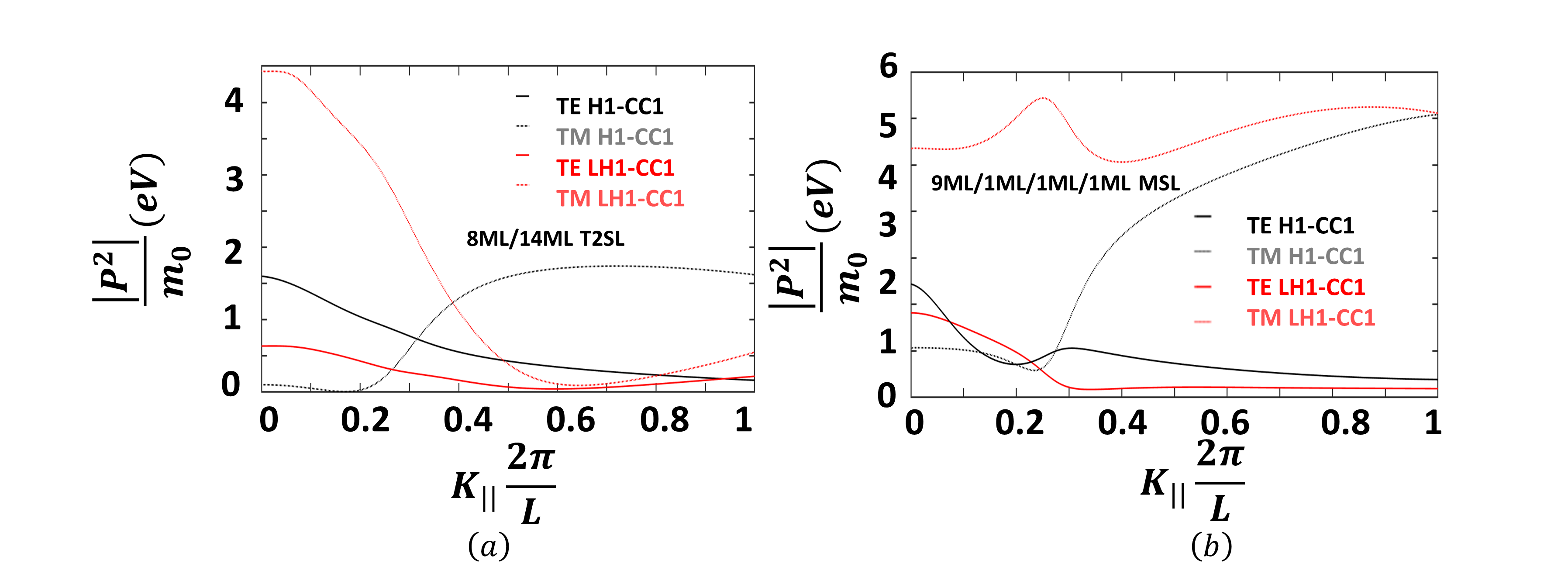}}
			\caption{Calculated Optical matrix element for the radiative transition between the various conduction and valence subbands considering the TE and TM polarization obtained by using Fermi's golden rule for T2SL and MSL in the in-plane direction. (a) TE(x) and TM(z) transition probabilities for  8ML/14ML T2SL (b) TE(x) and TM(z) transition probabilities for 9ML/1ML/1ML/1ML MSL. The dominant TE $C_{1}-HH_{1}$ transition increases as the $k_{||}$ increases and the TM $C_{1}-HH_{1}$ shows the opposite behavior. For the TE the $C_{1}-HH_{1}$ dominates but for the TM the C1-LH1 dominates.}
		\label{fig20}
\end{figure}
The squared envelope wavefunctions for the $C_{1}$, $HH_{1}$, $LH_{1}$, and $HH_{2}$ are plotted in Fig.\ref{fig18} (a) and (b) for T2SL and MSL, respectively. The high probability of finding holes in MSL is eminent in newly formed GaSb hole wells. Also, the center of $HH_{1}$ shifts further towards the interface, which results in an increased spatial product of envelope wavefunctions at the interfaces, as shown in Figs. \ref{fig19} (a) and (b). In this particular case of MSL, notice a higher $C_{1}$-$HH_{1}$ wavefunction overlap, as shown in Figs. \ref{fig19} (a) and (b) at the interfaces. Additionally, we calculate the oscillator strength of the $C_{1}$-$HH_{1}$ transition, which is higher for the MSL.\\
\begin{figure}[!htbp]	
	\quad
	\subfigure[]{\includegraphics[height=0.23\textwidth,width=0.45\textwidth]{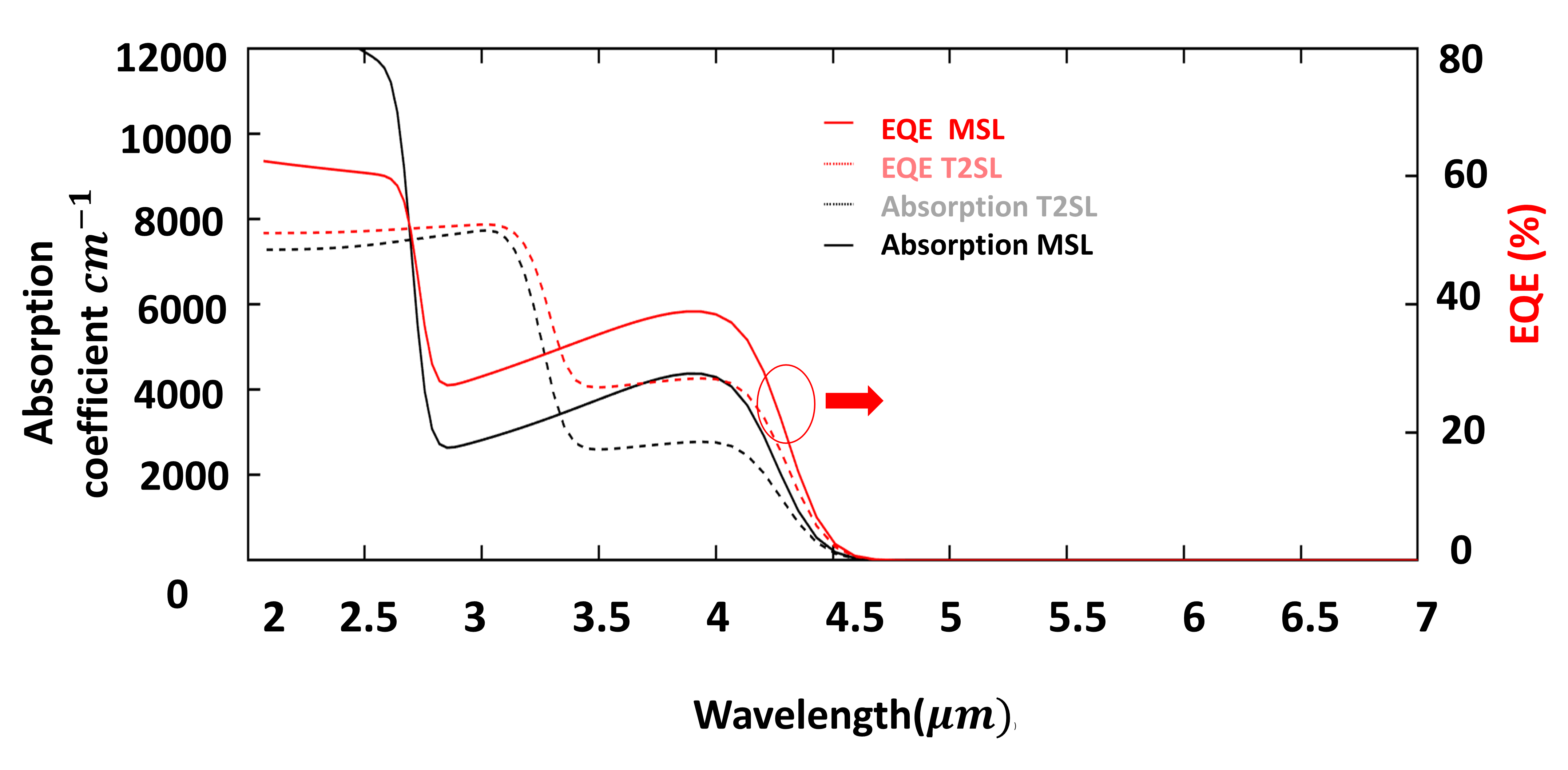}}
			\caption{Theoretical absorption spectra and external quantum efficiency for mid wavelength for 8ML/14ML T2SL and 9ML/1ML/1ML/1ML MSL.}
		\label{fig21}
\end{figure}
\begin{table}
    [!htbp]
\caption{Properties of superlattices calculated for the comparison of two same bandgap mid wavelength T2SL and MSL superlattices by the 8 band $\bf k.p$ perturbation at $T=77K$.} 
\label{table3}
\begin{center}      
\begin{tabular}{|l|l|l|l|l|}
\hline
\rule[-1ex]{0pt}{3.5ex}Parameters & T2SL& MSL\\
\hline
\rule[-1ex]{0pt}{3.5ex}Bandgap (eV) & 0.2892 & 0.2898 \\
\hline
\rule[-1ex]{0pt}{3.5ex}Electron effective mass & 0.0384 & 0.0298\\
\hline
\rule[-1ex]{0pt}{3.5ex}Hole effective mass & 0.1141& 0.0568\\
\hline
\rule[-1ex]{0pt}{3.5ex}Intrinsic carrier concentration $n_{i} (m^{-3})$  & $3.3x10^{12}$ & $1.52x10^{12}$\\  
\hline
\rule[-1ex]{0pt}{3.5ex}$B_{rad}(cm^{3}s^{-1}$) & $9.2x10^{-10}$ & $3.5x10^{-9}$ \\  
\hline
\end{tabular}
\end{center}
\end{table}
\begin{figure}[!htbp]
	\centering
		{\includegraphics[height=0.22\textwidth,width=0.45\textwidth]{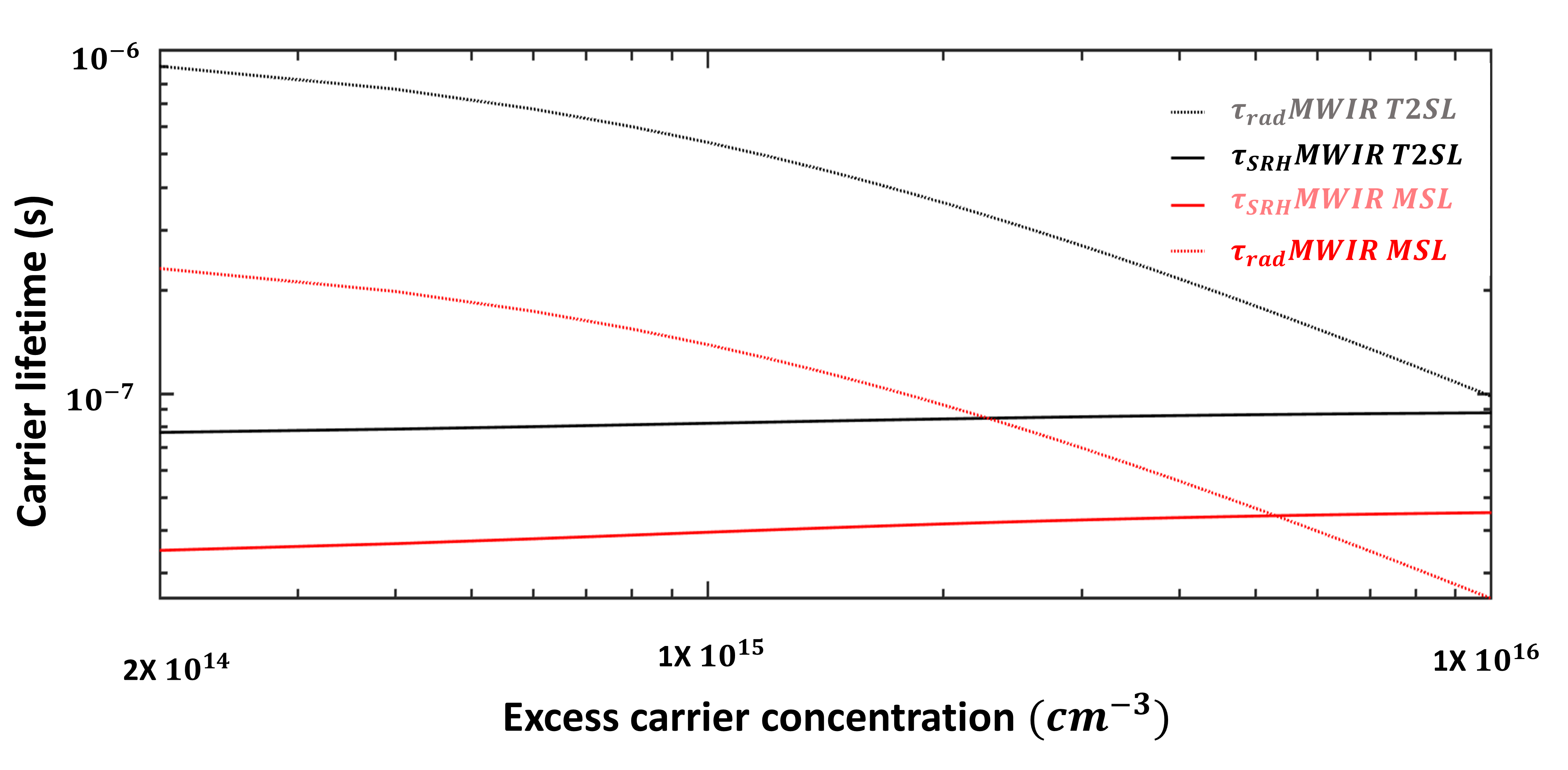}}
%\subfigure[]
		%{\includegraphics[height=0.23\textwidth,width=0.45\textwidth]{SRH_DN_MWIR.png}}
		\caption{ The radiative and SRH lifetimes calculated by the 8 band $\bf k.p$ method for the mid-wavelength T2SL and MSL when the excess carrier concentration is varied from $2x10^{14}-1x10^{16}cm^{-3}$. At low-level injection, the SRH recombination dominates the lifetime.}
		\label{fig22}
\end{figure}
\indent Next, in the Figs. \ref{fig20} (a) and (b), we plot the optical matrix elements for TE and TM polarization for further inspection of the optical transitions in the MWIR superlattices. At the lower $k_t$ values, the $C_{1}$-$HH_{1}$ TE components dominate, while the TM components are practically absent close to the Brillouin zone center \cite{wang2016enhancement}. The strongest behavior is seen by the TM components for the  $C_{1}$-$LH_{1}$ transition\cite{avrutin1993te}. Similar to what was mentioned for the SWIR absorbers, the TE and TM modes exhibit the opposite behavior. While TE for heavy holes and conduction band transitions reduces from higher to lower, TM increases from the zone center to the higher $k_t$\cite{wang2016enhancement}. Near the zone center, the TM component for the $C_{1}$-$LH_{1}$ is higher than the TE component.\\
\indent The calculated electron and hole effective masses,
ni and radiative recombination coefficients, are shown in Table. \ref{table3} for these superlattices.
In Fig. \ref{fig21} , we calculate the absorption coefficients for one period of both the T2SL and MSL and notice that the MSL has a higher absorption, which is related to its higher interface overlap and oscillator strength, as discussed above.   Further in Fig. \ref{fig22}, we plot the carrier lifetime mainly SRH and radiative for these superlattices, where we vary the excess carrier concentration from $2x10^{14}-1x10^{16}cm^{-3}$ and kept the absorber doping as $1x10^{15}cm^{-3}$. The obtained quantum efficiency for $2\mu m$ thickness of for T2SL and MSL absorber is 28\% and 38\%, respectively. Further in Fig. \ref{fig22}, we plot the carrier lifetime mainly SRH and radiative for these superlattices, where we vary the excess carrier concentration from $2x10^{14}-1x10^{16}cm^{-3}$ and kept the absorber doping as $1x10^{15}cm^{-3}$. We obtain the carrier lifetime around $39ns$ and $82ns$, at excess carrier concentration of $1x10^{15}cm^{-3}$, for MSL and T2SL, respectively. The ratio of diffusion dark current for $2\mu m$ of T2SL(303 periods) is twice of the diffusion dark of $2\mu m$ of MSL(556 periods). Next, we calculate the  quantum efficiency at the wavelength around $4 \mu m$ for $2\mu m$(thickness) of both of these superlattices, and we found the quantum efficiency for MSL and T2SL are 38\% and 28\%, respectively. Therefore, MSL shows better properties for MWIR  photodetection.
\begin{figure}[!htbp]	
\centering
%	\subfigure[]{\includegraphics[height=0.25\textwidth,width=0.45\textwidth]{Ek_8_6.png}\label{b1}}
{\includegraphics[height=0.25\textwidth,width=0.45\textwidth]{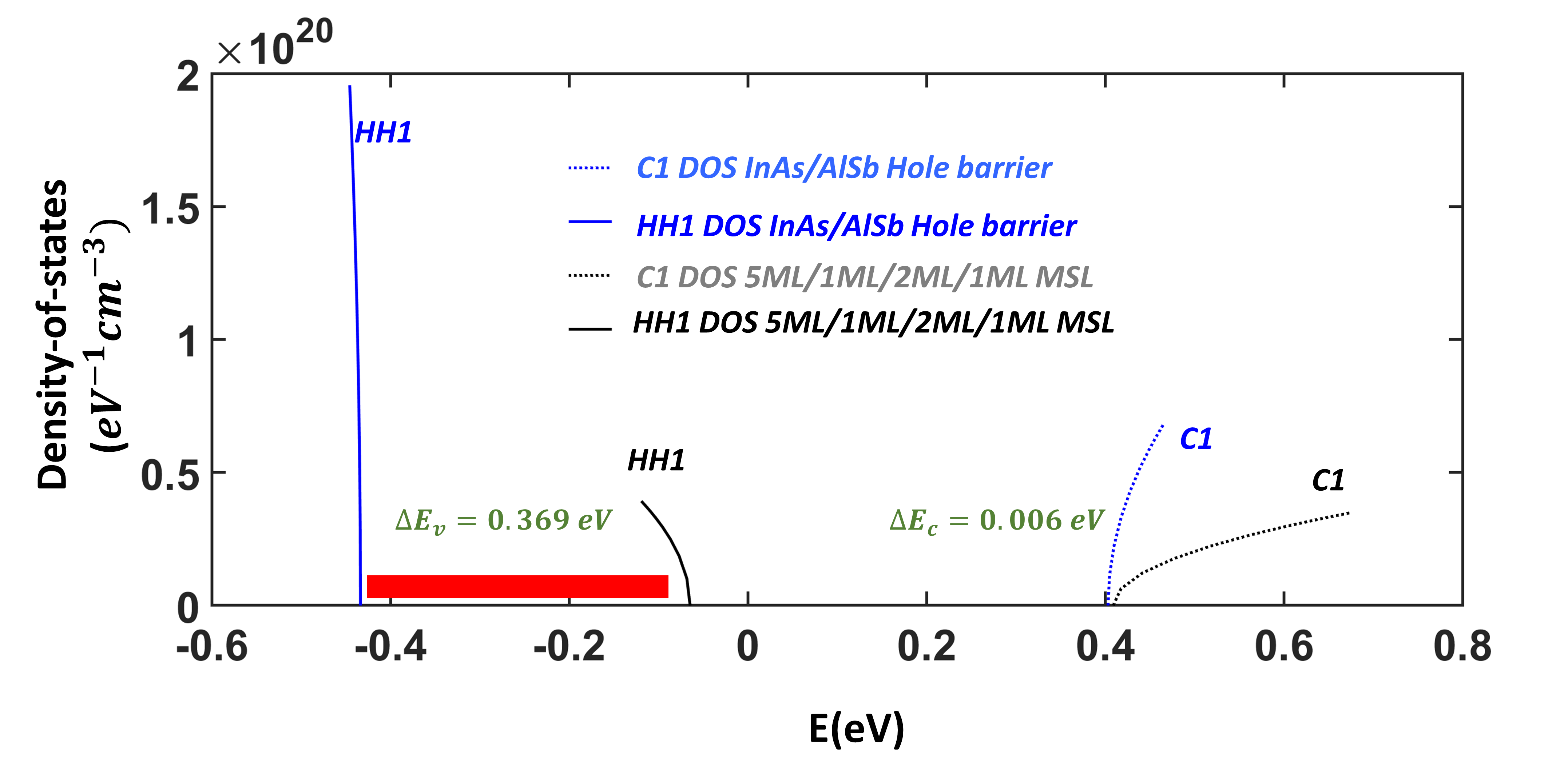}}
			\caption{Hole barrier designed for 5ML/1ML/2ML/1ML MSL by calculation of the band edges and density-of-states by the 8 band $\bf k.p$ method. The hole barrier is blocking the states from -0.043 eV to -0.43 eV(shown by red color) as it has no available states in this region.}
		\label{fig23}
\end{figure}
\subsection{Barrier}
An XBp photodetector consists of three regions: contact, barrier, and active or absorber region. Introducing a barrier in these detectors reduces the SRH and band-to-band tunneling currents; therefore, they perform better than traditional PIN detectors\cite{delmas2017design}. To restrict
the flow of carriers from the contact to the active region,
which may contribute to the thermionic currents. If not
appropriately treated, it may become equal to the diffusion current. Hence, there should be a proper band offset, which
prohibits this flow. Not just the band offsets, the doping
and the thickness of the barrier corresponding to the absorber
region play a crucial role in the dark current flow and carriers’ collection to their respective contacts.
Here, we plot the density-of-states(DOS) for the 6ML/7ML InAs/AlSb hole barrier and for the 5ML/1ML/2ML/1ML SWIR MSL absorber in Fig.\ref{fig23}. Here we can observe that the hole barrier provides a valence band offset of around 0.369$eV$ for the holes and is almost negligible i.e. 6$meV$ conduction band offset w.r.t to the absorber\cite{mir2013electrical}.
\section{Conclusion}
\label{conclu}
An exhaustive study of the InAs/GaSb type-II superlattice and InAs/GaSb/AlSb/GaSb M superlattice has been done to achieve high performance infrared photodetectors. The 8-band $\bf k.p$ theory, with the inclusion of the InSb/GaAs layer at the interface, was employed to calculate the band properties. The main results include bandgap variations, oscillator strength, and SRH capture lifetime constants. The absorption spectra of type-II lattices were calculated by utilizing Fermi's Golden rule. The absorption and photoluminescence were found to be functions of widths of the InAs and GaSb taken in one period of the superlattice. These calculations also led to the calculation of the radiative recombination coefficient, which was found to be higher for the lower widths of InAs and GaSb. Further, to increase the absorption and diffusion length in the type-II superlattice, we explored the M superlattice. We have studied the effects of introducing the high bandgap AlSb layer on the bandgap, carrier lifetime, and the possible reason for the red shift in the photoluminescence spectra. Based on the carrier localization and carrier spatial overlap at the interface, we have compared the same bandgap type-II and M superlattice for short and mid-wavelength photodetection. Given that the carriers at the interface overlap more in the short wavelength M superlattice than in the type-II superlattice, a greater quantum efficiency (with a comparable diffusion-dark current) has been predicted. On the other hand, the gain in quantum efficiency in M-superlattice has been seen with the reduction in the diffusion-dark current at the mid-wavelength.
 \section*{Acknowledgment}       
The authors acknowledge the funding from the PMRF Ph.D. scheme of the Ministry of Education, Government of India, and this work is also supported by ISRO-IIT Bombay Space Technology Cell.

\bibliography{reference}

\end{document}